\documentclass[3p,twocolumn]{elsarticle}%

\usepackage{lineno,hyperref}
\usepackage{subfig}
\usepackage{amssymb,adjustbox}
\DeclareMathSizes{10}{9}{7}{5} %All math texts at 9pt
\usepackage{relsize}
\usepackage{breqn}
\usepackage{multirow}
\usepackage{colortbl}
\modulolinenumbers[5]
\usepackage{graphicx}
\usepackage{tabularx}
\usepackage{amsmath}
\usepackage{booktabs}
\usepackage{color,soul}

\usepackage[version=4]{mhchem} 

\journal{Calphad (accepted for publication)}
\makeatletter
\def\ps@pprintTitle{%
  \let\@oddhead\@empty
  \let\@evenhead\@empty
  \def\@oddfoot{\reset@font\hfil\thepage\hfil}
  \let\@evenfoot\@oddfoot
}
\makeatother

\begin{document}

\begin{frontmatter}

%\title{Thermodynamic modeling of the FeTi-H system's para-equilibrium: an ab initio and experimental study}
\title{Modeling the thermodynamics of the FeTi hydrogenation under para-equilibrium: an \emph{ab-initio} and experimental study}
%\tnotetext[mytitlenote]{Fully documented templates are available in the elsarticle package on %\href{http://www.ctan.org/tex-archive/macros/latex/contrib/elsarticle}{CTAN}.}

\author[add1]{Ebert Alvares\corref{cor1}}
\ead{ebert.alvares@hereon.de}

\author[add1]{Paul Jerabek\corref{cor1}}
\ead{paul.jerabek@hereon.de}

\author[add1]{Yuanyuan Shang}

\author[add1]{Archa Santhosh}

\author[add1]{Claudio Pistidda}

\author[add2]{Tae Wook Heo}

\author[add3]{Bo Sundman}

\author[add1]{Martin Dornheim}

\address[add1]{Institute of Hydrogen Technology, Helmholtz-Zentrum Hereon, Max-Planck-Strasse 1, 21502 Geesthacht, Germany}

\address[add2]{Materials Science Division, Lawrence Livermore National Laboratory, Livermore, California 94550, USA}

\address[add3]{OpenCalphad, 9 Allée de l’Acerma, 91190 Gif sur Yvette, France}

\cortext[cor1]{Corresponding authors}

%\author{E. Alvares, Y. Shang, A. Santhosh, M. Dornheim, B. Sundman, P. Jerabek}
%\author{Ebert Alvares, Yuanyuan Shang, Archa Santhosh, Martin Dornheim, Bo Sundman, Paul Jerabek}
%\author{}
%\fnref{myfootnote}
%\address{Institute of Hydrogen Technology, Helmholtz-Zentrum Hereon, 21502 Geesthacht, Germany}
%\address{OpenCalphad, 9 Allée de l’Acerma, 91190 Gif sur Yvette, France}


%\author[mymainaddress,mysecondaryaddress]{Helmholtz-Zentrum Hereon}
%\ead[url]{ebert.alvares@hereon.de}

%\author[mysecondaryaddress]{Helmholtz-Zentrum Hereon}
%\corref{mycorrespondingauthor}
%\cortext[mycorrespondingauthor]{}
%\ead{ebert.alvares@hereon.de}

%\address[mymainaddress]{Max-Planck Strasse 1, Geesthacht}
%\address[mysecondaryaddress]{}

\begin{abstract}
FeTi-based hydrides have recently re-attracted attention as stationary hydrogen storage materials due to favorable reversibility, good sorption kinetics and relatively low costs compared to alternative intermetallic hydrides. Employing the OpenCalphad software, the thermodynamics of the (FeTi)$_{1-x}$H$_{x}$ (0 $\leq x \leq$ 1) system were assessed as a key basis for modeling hydrogenation of FeTi-based alloys. New thermodynamic data were acquired from our experimental pressure-composition-isotherm (PCI) curves, as well as first-principles calculations utilizing density functional theory (DFT). The thermodynamic phase models were carefully selected based on critical analysis of literature information and \emph{ab-initio} investigations. Key thermodynamic properties such as dissociation pressure, formation enthalpies and phase diagrams were calculated in good agreement to our performed experiments and literature-reported data. This work provides an initial perspective, which can be extended to account for higher-order thermodynamic assessments and subsequently enables the design of novel FeTi-based hydrides. In addition, the assessed thermodynamic data can serve as key inputs for kinetic models and hydride microstructure simulations.
\end{abstract}


\end{frontmatter}

% \linenumbers

\section{Introduction}

Binary intermetallic compounds with the general formula AB$_{n}$ carry envisaged thermodynamic and kinetic properties for a variety of hydrogen storage scenarios \cite{Hirscher2010}. Elemental hydrides of early (A) and late (B) transition metals generally possess high negative and positive formation enthalpies ($\Delta$H$_{f}$), respectively \cite{Dematteis2021a}. However, it is often observed that combining A and B in certain ratios (e.g. $n = 1, 2, 3, 5$) promotes the formation of stoichiometric compounds AB$_{n}$  together with a reduction of $\Delta$H$_{f}$ of their correspondent ternary hydrides \cite{Lys2020}. This reduction can improve hydrogenation reversibility in scenarios close to ambient conditions, and consequently provide a safer operational hydrogen storage medium when compared to molecular state hydrogen tanks \cite{Dematteis2021a,Lys2020}.

Moreover, these intermetallics exhibit large variability of the thermodynamics
and kinetics of hydrogenation when partially substituting other suitable metallic elements for their base components.

% This indicates the possibility of engineering these compounds in order to tune their performance for the desired applications \cite{Lys2020}.

%Moreover, the thermodynamics and kinetics of hydrogenation of these intermetallics can be tuned by partial chemical substitution of their base components by other suitable metallic elements, which makes these compounds a flexible class of materials that allows fine adjustment of their performance tailor-made for the desired applications \cite{Lys2020}.

Among these compounds, FeTi is recognized as a key material due to the possibility of large-scale production owing to its relatively low cost compared to other intermetallics.

%Besides, its application as solid-state hydrogen storage material potentially poses no significant complications for re-using and recycling as far as we know. However, this needs to be further investigated and is therefore an area of active research for some of us.

FeTi has gravimetric and vo\-lu\-me\-tric capacity of 1.87 wt.~\% and 105~kg$_{\mathrm{H}_{2}}$/$\mathrm{m}^3$ \cite{Dematteis2021a}, respectively, and combines good sorption kinetics and reversibility within operating conditions near room temperature and atmospheric pressure. The reversible volumetric capacity of FeTi (83.7~kg$_{\mathrm{H}_{2}}$/$\mathrm{m}^3$) \cite{Broom2011} as well as many other metal-hydrides (MHs) even surpasses cryogenically liquefied hydrogen (71.42~kg$_{\mathrm{H}_{2}}$/$\mathrm{m}^3$ at 20~K) \cite{Sasaki2016}. Therefore, they represent an excellent storage option when weight of the system is not a concern, e.g. hydrogen supply for residential environments, emergency power supply and for heavy-weight means of transportation like trains, ships, long-haul trucks etc. \cite{HyCARE2019, GKN2021}.

%Although FeTi is an attractive hydrogen storage medium for these reasons, some obstacles regarding its activation and kinetics still hinder widespread application.

Despite these promising properties as a hydrogen storage medium, the widespread adoption of FeTi has been hampered by difficulties in initial activation for hydrogenation. As recently reviewed by Dematteis et al. \cite{Dematteis2021a}, this issue may be overcome by partial chemical substitution of the alloying elements. The authors provided the lists of substituting elements for improving properties required for in-service performance, as well as elements to avoid for economical or environmental reasons. They indicated the importance of in-depth systematic studies to elucidate synergistic effects of multicomponent substitution on the thermo-kinetic properties of FeTi-based hydrides. However, comprehensive studies and understanding of the associated mechanisms are still noticeably lacking. 

In this context, the Calphad (Calculation of phase diagrams) method has been recognized as an effective tool to predict thermodynamic properties of multicomponent materials solely based on descriptions of their lower-level unaries, binaries and ternaries that make up the higher-order system \cite{Kattner2016,Lukas2007}. Therefore, the development and/or improvement of MHs systems for hydrogen storage may greatly benefit from computational thermodynamics at the design and simulation stages, since it offers excellent flexibility and significant time savings. In particular, the computational thermodynamic approach allows us to interrogate the thermodynamic properties of new materials with varying chemistry \textit{in-silico} before performing real-life synthesis and testing in the laboratory.

The present work aims at modeling the thermodynamics of the FeTi alloy hydrogenation focusing on its structural and thermodynamic features. The reported crystal structures of the relevant phases were used as a starting point for the thermodynamic modeling as well as to guide \emph{ab-initio} calculations of thermochemical properties of FeTi hydrides. Newly measured experimental pressure-composition-isotherm (PCI) curves supplemented the available information on hydrogen solubilities. The literature review and the acquired data in this work provided the key information to perform the Calphad assessment. The calculated phase diagrams and thermodynamic properties are discussed and compared with the reported data.

\section{Literature review} %on the FeTiH$_{x}$ system

In this section, we review the relevant thermodynamic and crystallographic properties of MHs and FeTi-hydrides that informed the thermodynamic modeling performed in the present work.

In subsection \ref{fetialloysec}, important properties of Fe and Ti for hydride formation are presented. In subsection \ref{fetihydridesec}, crystallographic features of the FeTi-hydrides are reviewed. Subsection \ref{para-eqsec} introduces the para-equilibrium concept as a basis for multi-component hydrides formation. Finally, in subsections \ref{hydrogenationsec} and \ref{fetiphdiasec}, the hydrogenation process and equilibrium phase diagrams of FeTi-H system are revisited. Subsection \ref{implicationssec} contains summarizing remarks and implications for the thermodynamic model.

\subsection{FeTi alloy}\label{fetialloysec}

Both Fe and Ti metals show a temperature range in which the bcc structure is stable. Nonetheless, they have a distinguished affinity to hydrogen and, therefore, different behavior upon hydrogenation. The calculated hydrogen enthalpy of solution in the bcc lattice of iron is 24.12~kJ/mol$\cdot$at\footnote{Note that the utilized unit for the enthalpy (kJ/mol$\cdot$at) refers to a mole of \textit{any} element in the respective formula, not only hydrogen (moles of H or H\textsubscript{2}) as sometimes encountered in literature about thermodynamics of metal hydride systems.}, whereas in the bcc lattice of titanium it is -59.82~kJ/mol$\cdot$at \cite{Fukai2005}. 

Although hydrogen diminishes the temperature range where bcc iron is stable, it widens the bcc solid solution field in the Ti-H phase diagram. Herewith, iron hydride (FeH) forms only under severe hydrogen pressure, above 6~GPa at 523~K \cite{Zinkevich2002,Antonov1981}, yet titanium hydride (TiH$_{2}$) is very stable and can form even below atmospheric pressure ($<$~10$^{5}$~Pa) at 576~K \cite{Wille1981,San-Martin1987,Arita1982}.
 
The absorption of hydrogen in solid metals is known to occur interstitially, especially for the bcc metals hydrogen preferentially occupies tetrahedral sites \cite{Thompson1980}. Iron and titanium follow this trend, as evidenced by crystallographic and thermodynamic analyses   \cite{Fukai2005,DaSilva1976,Jiang2004,Wille1981,Beck1975,San-Martin1987}. 

When Fe and Ti mix in equimolar quantities, the system's energy reduces by ordering the unlike atoms in a bcc-like structure where the Ti atoms position in the corners and the Fe atom in the center of the cube, as the interposition of two primitive cubic sublattices of Fe and Ti (CsCl-type structure).  
 
The ordering of the metallic elements gives rise to two distinguished configurations of octahedral interstices, Ti$_{4}$Fe$_{2}$ and Ti$_{2}$Fe$_{4}$, with Fe and Ti as nearest neighbors, respectively. Because Ti has a larger atomic radius compared to Fe, the Ti$_{4}$Fe$_{2}$ interstices ($\approx$ 0.31~\r{A}) are noticeably enlarged compared to Ti$_{2}$Fe$_{4}$ ($\approx$ 0.09~\r{A}) \cite{Thompson1980}.  
 
\subsection{FeTi-hydrides}\label{fetihydridesec} 
 
\begin{figure}[htp]
    \centering
    \includegraphics[width=6cm]{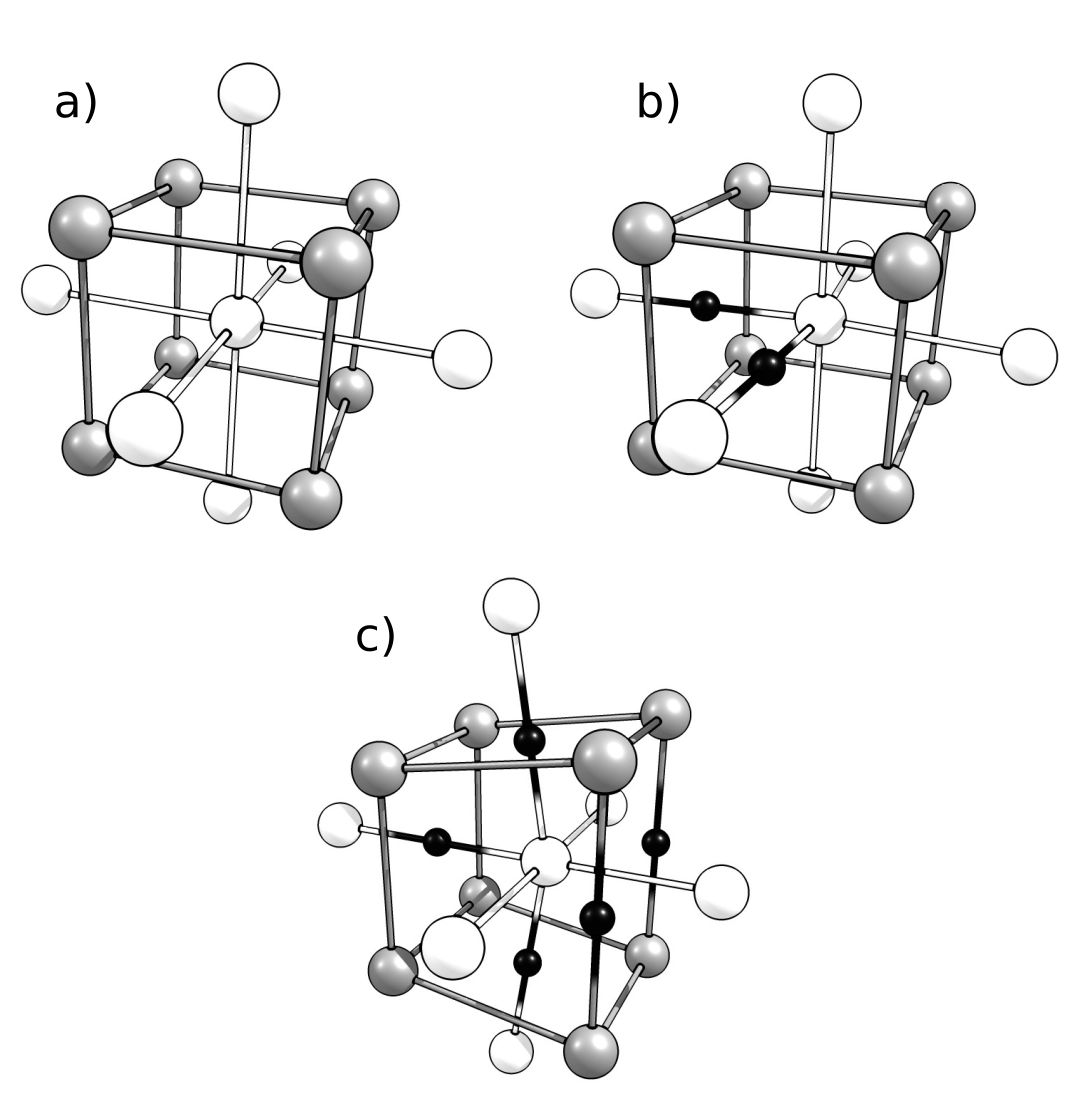}
    \caption{Example of ternary hydride structures from the bcc-like unit cell framework including first neighbors iron atoms. a) B2 (purely metallic FeTi), b) $\beta$ (monohydride) and c) $\gamma$ (dihydride). Ti: grey, Fe: white, H: black.  }
    \label{fig:Structures}
\end{figure}

Reilly and Wiswall were the first to study the hydriding properties of FeTi  \cite{Reilly1974}. Hydrogen absorption in FeTi yields ternary hydrides and is associated with symmetry reduction corresponding to considerable distortions of the CsCl-type metal structure. At room temperature, the FeTi-hydrogenation leads to the solid solution region ($\alpha$) up to the composition FeTiH$_{x \approx 0.1}$, the monohydride
FeTiH$_{1.04  \mathsmaller\lessapprox  x  \mathsmaller\lessapprox  1.22}$ ($\beta$) and the dihydride FeTiH$_{1.71  \mathsmaller\lessapprox  x  \mathsmaller\lessapprox  1.93}$ ($\gamma$). Examples of each phase from the perspective of the bcc-like frame are shown in Fig. \ref{fig:Structures}.

For each FeTi-hydride phase, neutron diffraction (ND) experiments have been performed and indicate that hydrogen atoms occupy octahedral interstices \cite{Thompson1978,Thompson1979,Thompson1980,Schober1979,Schafer1980, Fischer1978}. 
Transmission electron microscopy (TEM) and X-ray diffraction (XRD) analysis show that the FeTi-H system is isostructural with FeTi-D \cite{Schober1979,Schober1980,Schafer1980,Reidinger1982}, i.e. the isotopic effect is negligible concerning crystal structure and thus it can be assumed that there is a corresponding hydride for each deuteride. For this reason, in the next subsections we refer to hydrides even though some structural experiments were carried out using deuterides.\\ 

In the following we compile the experimental findings for the $\alpha$-, $\beta$- and $\gamma$-hydrides that are the basis for our thermodynamic model.

\subsubsection{$\alpha$-hydride}
The $\alpha$ hydride forms via hydrogen dissolution in the interstitial sites of the CsCl-cubic lattice with H being more likely to enter into the octahedral sites of the B2 phase (Fig. \ref{fig:Structures} a). Density functional theory (DFT) calculations \cite{Nong2014} demonstrated that the Ti$_{4}$Fe$_{2}$ octahedral configuration has a more negative hydrogen absorption energy compared to Ti$_{2}$Fe$_{4}$ and thus is more likely to host a hydrogen atom, which agrees with ND experiments \cite{Thompson1980}. The order parameter ($Q$) indicates the ordering of Fe and Ti in the bcc-like lattice. A perfect ordered crystal has $Q$ calculated to be unity. Fischer measured $Q$ of the $\alpha$ phase and found that the order parameter reduction diminishes only slightly even after several cycles of hydrogen sorption and corresponds mostly to surface effects \cite{Fischer1978}, which indicates that the $\alpha$ hydride has an ordered metal structure analogous to the FeTi-B2 phase with enlarged lattice para\-meter (a$_{\alpha}$ = 2.979~\r{A}) and H randomly occupying the available Ti$_{4}$Fe$_{2}$ octahedral sites \cite{Thompson1980,Schafer1980}. 

\subsubsection{$\beta$-hydride}
Additional hydrogen insertion increases the displacement of nearest Fe atoms causing anisotropic variation on lattice parameters and leading to symmetry reduction. As consequence,  $\beta$ has an orthorhombic structure \cite{Thompson1978} with space group P222$_{1}$. The indexation of the $\beta$ phase finds lattice parameter $b_{\beta} \approx $ 4.3~\r{A}, corresponding to the diagonal of the CsCl cube of FeTi, as well as $a_{\beta}\approx$ 4.5~\r{A} and $c_{\beta} \approx $ 2.98~\r{A}. Two Ti$_{4}$Fe$_{2}$ octahedral sites, H1 and H2, can be identified in the $\beta$ structure, which are partially occupied.

The $\beta$ hydride presents two distinguished forms, $\beta_{1}$ and $\beta_{2}$ \cite{Schober1979,Schefer1979,Reidinger1982}. They have a similar structure, except for the hydrogen distribution in the H1 and H2 sites. Schäfer et al. \cite{Schafer1979} discussed this for the cases when the interstitial atoms are evenly occupied, i.e. H1/H2 = 0.50/0.50, or distributed in a 0.90/0.10 ratio.  
Despite this, experimental investigations show an ordered hydrogen occupation in the sublattices of both forms of $\beta$ \cite{Thompson1978, Schefer1979, Schober1980}. 

The occupation ratio of $\beta_{1}$ was observed to be 0.88/0.12 \cite{Schefer1979, Schafer1979}, while $\beta_{2}$ has it as 0.92/0.45,  corresponding roughly to the FeTiH and (FeTi)$_{3}$H$_{4}$ formula units, respectively.

The volume expansion of $\beta$ upon hydrogenation reaches 4.5\% with experimental evidence suggesting that the $\beta_{1}$~$\rightleftharpoons$~$\beta_{2}$ transition is a discontinuous transformation at room temperature. However, to a certain extent it might actually be rather described as a continuous single-phase process with both sites simultaneously being enriched with hydrogen \cite{Schober1979, Schefer1979}.

\subsubsection{$\gamma$-hydride}
The analysis of the $\gamma$ hydride crystal structure ($x >$~1.74) is still under debate due to its demanding synthesis that requires hydrogen pressures above 5~MPa. Moreover, the presence of strain and remaining amounts of $\beta$ phase complicate an in-depth study even further.
 
After its first X-ray diffraction detection suggesting a cubic structure \cite{Reilly1974}, Fischer et al. reported that ND indicated rather an orthorhombic structure \cite{Fischer1978}. Shortly after, ND and TEM studies correlated the $\gamma$ structure with the orthorhombic $\beta$ \cite{Schefer1979, Thompson1979}, leading to conclusions that $\gamma$ forms from the full insertion of hydrogen in the previously observed Ti$_{4}$Fe$_{2}$ sites of $\beta$ and in an additional Ti$_{2}$Fe$_{4}$ octahedral site, pushing towards the formation of a monoclinic P2/m structure, losing orthogonality and degrees of symmetry.

Schefer observed that the analogous hydride compound also shows such a transformation \cite{Schefer1979} and Schäfer et al. \cite{Schafer1980} identified an increase of mono\-cli\-nic distortion of $\gamma$ over decreasing temperature.

Fischer et al. \cite{Fischer1987} concluded that the Ti$_{2}$Fe$_{4}$ sites are energetically much less favourable and possibly the last to be filled by hydrogen, reaching a partial occupation with 91\% at large hydrogen concentrations. It was assumed that the orthorhombic Cmmm structure better suits ND data than the mono\-cli\-nic P2/m, agreeing with later observations by Thompson et al.  \cite{Thompson1989} and DFT calculations of the fully hydrogenated  $\gamma$ (FeTiH$_{2}$).\\

\subsection{Calphad modeling of MHs \& para-equilibrium}\label{para-eqsec}

Special care should be taken when Calphad modeling is applied to multicomponent MHs because peculiar thermodynamic features are often present in these systems. As metallic atoms (M) significantly differ from hydrogen atoms (H), especially in mass and size, M atoms establish the main lattice of the crystal structure while H atoms enter the bulk and partially occupy the interstitial sites. 

In stark contrast to the low mobility of metal atoms, hydrogen atoms have a much higher mobility at room temperature. As a consequence, H responds to its chemical potential ($\mu_{_{\mathrm{H}}}$) by quickly diffusing into the interstitial sublattice forming hydrides. 

When this situation is attended, the hydrogen chemical potential equalizes in each of the co-existing phases, while this doesn't happen for the metallic elements that keep the alloy composition constant. Reaching complete equilibrium is hindered by the sluggish diffusion kinetics of metallic elements, causing a meta-stable thermodynamic equilibrium of the phases. Such an internal meta-stable state is considered a para-equilibrium.\footnote{Flanagan and Oates \cite{Flanagan2005} reviewed the various degrees of equilibria related to two-phase co-existence regions of M-H systems to which we refer for a more detailed explanation.} Under this perspective, the Calphad framework have been applied to model a number of metal-hydrogen systems. Joubert \cite{Joubert2012} comprehensively reviewed the application of the Calphad method to model MHs and documented the main challenges he encountered.\\

As mentioned in subsection \ref{fetihydridesec} and will be seen later in the text, the higher stability of TiH$_{2}$ in relation to the FeTi ternary hydrides ensures that the hydrogenation of FeTi can be considered as a para-equilibrium process. Subsequently, the analysis of the phases' stability and the application of the Calphad framework to model the para-equilibrium of the FeTi-H system in this work is presented.

\subsection{FeTi-hydrogenation}\label{hydrogenationsec}

The \emph{in-situ} investigation of the hydrogen storage properties for hydride systems is generally performed at different isothermal cycles of repeated absorption and desorption by carefully  equilibrating hydrogen gas with the metallic sample in small pressure increments utilizing a Sieverts-type apparatus. The trace of the pressure over hydrogen sorption during the cycles results in so-called pressure-composition-isotherm (PCI) curves that are used for material characterization regarding temperature dependent hydrogen absorption/release. With increasing H\textsubscript{2} pressure, the hydrogen content in the solid system grows. Inflections of the PCI curves indicate the formation of new phases and, in an ideal scenario, the co-existence of two solid phases results in a completely flat plateau pressure. The initial and final points of the plateau correspond to the onset and termination of the solid phase transformations and indicate their equilibrium composition. 
Since the stability of solid phases has negligible dependence on pressure, a temperature-composition phase diagram of metal hydrides correlates to the solubility limits extracted from PCI curves.

Many authors studied the hydrogenation of the FeTi intermetallic \cite{Tessier1995, Reilly1974, Wenzl1980, Yamanaka1975}. Recently, Sujan et al. reviewed FeTi hydrogenation and processing, whereas Dematteis et al. reviewed the substitutional effects on hydrogen storage properties of FeTi \cite{Sujan2020, Dematteis2021a}.
 
For FeTi, there is a consensus that the hydrogenation around ambient temperatures involves at least three hydride phases, $\alpha$, $\beta$, and $\gamma$. Initially, hydrogen is dissolved in the ordered B2 phase forming the solid solution $\alpha$ phase. The increase in H$_{2}$ pressure leads to the saturation of $\alpha$ after which the external thermodynamic conditions maintain while hydrogen is absorbed, forming a new phase $\beta$. This phenomenon corresponds to the co-existence of these phases in equilibrium with the gas.

The completion of the H$_{2}$ + $\alpha$ $\rightleftharpoons$ $\beta$ transformation accompanies the cessation of hydrogen absorption at constant pressure. At this point, with constant temperature hydrogen is further absorbed only by incrementing its pressure. The precise solubility range of $\beta$ and the mechanisms of transformations it undergoes upon hydrogenation are not yet fully understood but seem to be highly dependent on the sample history, temperature and sorption path  \cite{Reidinger1982}.
 
After complete formation of $\beta$, at lower temperatures the increment in hydrogen pressure results in a smooth inflection of the pressure-composition curve, particularly visible in the logarithmic scale as a slopped plateau. 
At higher temperatures, the slope increases and the inflections become less evident. Above a critical temperature, there is no clear evidence of this phenomenon, and the exact point is difficult to determine since the transition is narrow and quite gradual.

%Since there are temperature and composition hysteresis and evidence that the initial state %and history of the sample highly influence the behavior of transitions occurring after the %formation of $\beta$, the mechanism of hydrogenation in this region is still unclear.

The formation of the $\gamma$ phase occurs at maximal hydrogenation, and experimental evidence shows that it may be a highly stable structure in the system, whilst the formation of $\beta_{2}$ is possibly only a metastable process. This is for example indicated by Reidinger et al. \cite{Reidinger1982} by the fact that annealing at 800~°C followed by activation facilitates the formation of $\gamma$, which could then be detected in coexistence with $\alpha$ at FeTiH$_{1.12}$. Another hint is the observations by Goodell et al. \cite{Goodell1980} showing that by increasing the number of sorption cycles, the second plateau fades out and higher hydrogen pressure is required to reach the same level of hydrogen absorption, although the first plateau pressure remains unaffected.

Both pieces of evidence give rise to the possibility that once $\gamma$ forms, it remains in the system, blocking the formation of the intermediate $\beta_{2}$ phase when a certain threshold amount of $\gamma$ is present.

We want to note that even though corresponding deuterides and hydrides in the FeTi-H/D systems are isomorphic \cite{Schafer1980}, their lattice parameters \cite{Schober1980} and the equilibrium pressure \cite{Fischer1978} slightly vary. Some transition points observed in deuterides might slightly shift in comparison to corresponding hydrides and thus our hypotheses have to be taken with a grain of salt.

\subsection{The FeTi-H phase diagram}\label{fetiphdiasec}

\begin{figure}[htp]
    \centering
    \includegraphics[width=8cm]{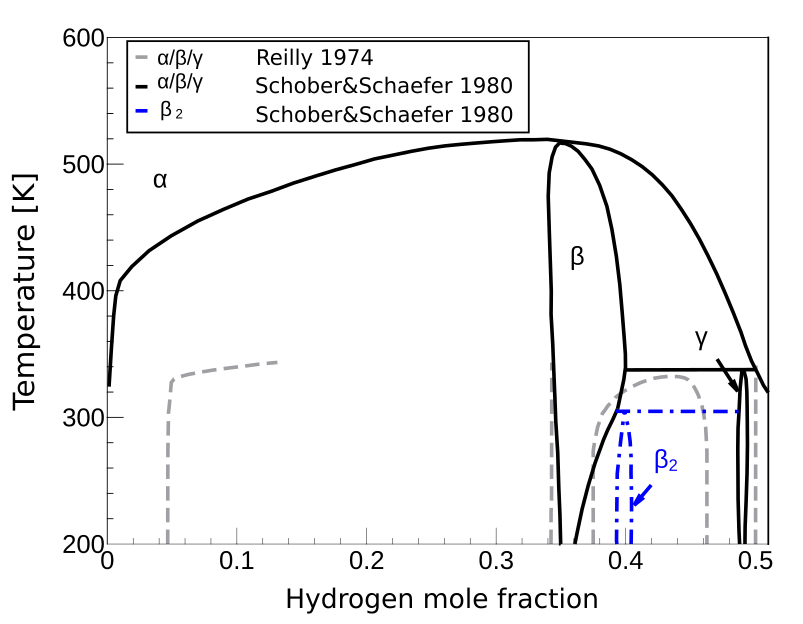}
    \caption{Adaptation of the schematic FeTi-H phase diagram from Schober and Schaefer \cite{Schober1980} superimposed to the one proposed by Reilly and Wiswall \cite{Reilly1974}. See text for detailed explanation.}
    \label{fig:literature-phdia}
\end{figure}

The first proposed phase diagram for the FeTiH$_{x}$ system \cite{Reilly1974} indicates that the solubility limit of $\alpha$ and the equilibrium composition of $\beta$ with $\alpha$ would be constant at least up to 70~°C. It also shows the existence of a critical temperature between 55~°C and 70~°C above which a discrete coexistence of $\beta$ and $\gamma$ can no longer occur and the transformation should become continuous (indicated by dashed lines in Fig.~\ref{fig:literature-phdia}). This phenomenon was assumed based on the lack of evidence of a second plateau for the 70~°C isotherm curve and the vestigial plateau trace at 55~°C.

Later, Bowman et al. \cite{BOWMAN197897} extended the phase diagram by compiling higher-temperature PCI sets from Reilly \cite{Reilly1974, Reilly1976} and Yamanaka \cite{Yamanaka1975}. Despite some uncertainties, there was a good agreement with respect to $\alpha$ to $\beta$ phase limits. Bowman's NMR study revised the FeTiH$_{x}$ diagram (including some range of uncertainty for low-temperature lines), extended the $\alpha$+$\beta$ phase limits to a higher temperature and showed an absence of the critical temperature for the $\beta$+$\gamma$ two-phase region below 127~°C \cite{BOWMAN197897}. He also suggested a $\gamma$-$\beta$ transformation at 117~°C either because of a critical stability temperature or due to some release of hydrogen.

In light of calorimetric and structural data in the range between 1--72~°C,  Lebsanft \cite{Lebsanft1978} proposed a variation of the $\alpha$-$\beta$ solubility limit, indicating that somehow they would converge to a critical point at higher temperatures and that the limit of stability of $\beta$ should be much closer to FeTiH$_{2}$, showing that $\gamma$ is possibly only stable as a quasi stoichiometric compound. Even though not mentioned by Lebsanft in his thesis, Schefer et al. \cite{Schefer1979} claimed that an additional plateau was visible in the low-temperature curves from Lebsanft and thus that the $\beta_{2}$ phase could be forming.

Soon after, Schober analyzed the system by TEM at varying temperatures. Cooling of the samples confirmed that $\beta$ and $\gamma$ could be kept in cryogenic temperatures, while sample heating revealed a peritectoid transition $\gamma$ $\rightleftharpoons$ $\alpha$ + $\beta$ above 60~°C \cite{Schober1979}. An additional phenomenon was later observed by Schober and Schäfer upon cooling under 35~°C \cite{Schober1980}. They found that plate-like phases precipitate in hydrogen-rich $\beta$ regions, agreeing with the previous evidence \cite{Schefer1979} of an existing ordered $\beta_{2}$ in equilibrium with $\beta_{1}$. Their published schematic phase diagram suggested another peritectoid reaction $\beta_{2}$ $\rightleftharpoons$ $\gamma$ + $\beta_{1}$ around 35~°C at FeTiH$_{1.33}$, as shown by dash-point lines in Fig.~\ref{fig:literature-phdia}.

More recently, Endo et al.~\cite{Endo2013} used \emph{in-situ} synchrotron radiation X-ray diffraction (SR-XRD) analysis under severe hydrogen fluid pressure of 5~GPa at 600~°C and observed a hydrogen-induced order-disorder transition from the  $\alpha$-like ordered FeTi solid solution into a bcc hydride. Under these harsh conditions, the $\alpha$-like phase presented higher hydrogen solubility and transformed into the bcc hydride, which underwent a lattice expansion of 21\% at its maximal hydrogenation stage. After some time, the bcc hydride decomposed into TiH$_{2}$ and Fe$_{2}$Ti.

We conclude this subsection by noting that the FeTiH$_{x}$ phase diagram is essentially metastable since the hydrogenation of FeTi occurs under para-equilibrium conditions. Thus, there is no confirmation that the kept metallic equimolar quantity of the hydrides in the two-phase regions is in their thermodynamically most stable form. However, since the alloy keeps its composition due to the sluggish diffusion kinetics of metal elements at lower temperatures and because the enthalpies of formation of TiH$_{2}$ and Fe$_{2}$Ti are much lower compared to those of the ternary hydrides \cite{Yamanaka1975}, the system tends to disproportionate to eventually reach complete equilibrium, as observed by Endo et al. \cite{Endo2013}.

\subsection{Implications for the thermodynamic model}\label{implicationssec}
We finish the literature review of the FeTi hydrogenation by concluding that from a technical perspective, the $\alpha \leftrightarrow \beta$ transformation appears to have the highest relevance for hydrogen storage applications and therefore will be the core of our thermodynamic model presented in this work. Nevertheless, in the future the $\gamma$-hydride should to be taken into account as well due to its role in limiting reversibility of the (de)hydrogenation processes.

\section{Experimental \& computational methods}

\subsection{Measurement of PCI curves}

Fe foil (99.5\% purity, purchased from Sigma-Aldrich) and Ti foil (99.7\% purity, purchased from Alfa Aesar) were placed into an arc melter and samples with a total of 4~g were melted together. The resulting ingot with an atomic ratio of 1:1 was re-melted at least five times to ensure homogeneity, and then suction-cast into a rod-shaped copper mold with dimensions of  3~mm diameter  and 30~mm length. 

The FeTi rod was transferred into  an argon-filled glovebox, with a circulation purifier controller to keep  O${_2}$ and H${_2}$O levels lower than 1~ppm to minimize oxidation and degradation of the material. The alloy was hand crushed and then sifted through a 125~$\mu$m sieve. The fine powder material was activated at 90~°C under dynamic va\-cuum for 15~h, after that 10 cycles were used to further activate the sample. The cycling conditions were 50~bar and 50~°C for absorption followed by 1~bar and 50~°C for desorption. A sample of approximately 1.8~g was used for the PCI measurements. 

PCI curves were recorded at 40, 50, 60, 70, and 100~°C under ultra-pure hydrogen gas atmosphere using the Sieverts method (employing Setaram PCT-Pro 2000 \& Hy-Energy) \cite{Sieverts1907}. The pressure was controlled so that an increment step was performed when the absorbed hydrogen content did not change within 10~min in mono-phase regions. For the points near and on the plateau, the waiting time at constant pressure was about 120~min for absorption and 180~min for desorption to allow enough time for equilibration.

\subsection{First-principles calculations}

 Spin-polarized DFT calculations for the known and hypothetical bulk phases in the pseudo-binary FeTi-H system were carried out with  the Vienna \textit{ab-initio} simulation package (VASP) version 5.4.4  \cite{Kresse1993, Kresse1996a,Kresse1996, Kresse1999}. For the description of exchange and correlation effects, the revised version of the Perdew-Burke-Ernzerhof functional for solids and surfaces (PBEsol) \cite{Perdew2008} within the generalized gradient approximation (GGA) was used. 
 
 Long-range dispersive (van der Waals) interactions were incorporated with the DFT-D3 correction method including Becke-Johnson damping \cite{Grimme2010, Grimme2011}. We made use of the projector augmented-wave method (PAW) \cite{Blochl1994} implemented in VASP and included all plane waves within a kinetic energy cut-off of 450 eV for the calculations. 
 
 All structures were fully optimized (\texttt{ISIF=2}) until the residual atomic forces were less than 0.01~eV$\cdot$\AA$^{-1}$ with an energy tolerance of 1$\times$10$^{-5}$~eV. A Monkhorst-Pack scheme \textbf{k}-point spacing of 0.4~\AA$^{-1}$ was used for the calculations. In order to account for the strong electronic correlations of the localized orbitals, an on-site Hubbard U correction \cite{Anisimov1991, Dudarev1998, Bengone2000} of 4.0 and 4.3~eV was applied to Ti and Fe, respectively. 
 
 The total energy of hydrogen (necessary to compute the respective formation energies) was obtained at the same level of theory by placing H\textsubscript{2} in a large unit cell with sufficiently large vacuum space around the molecule to avoid self-interactions caused by the periodic boundary conditions.

\subsection{Optimization of thermodynamic parameters}
The optimization of the thermodynamic model parameters was carried out with the OpenCalphad software version 6.25 \cite{Sundman2015}. The data relating solubility limits over temperature were taken from the measured PCI curves. The calculated DFT ground-state formation energies of each end-member of the phase models were assumed to be a reasonable approximation to the enthalpy of formation at room temperature (298.15~K) and therefore employed as experiments during the optimization of the model parameters. 
For the $\alpha$ phase solubility limits, the experimental data came from absorption curves where $\alpha$ is in equilibrium with the H$_2$ reservoir.

Neither absorption nor the desorption pressure-composition curves represent the true equilibrium \cite{Flanagan2005}. Despite the high level of hysteresis, the average pressure between absorption and desorption plateaux were used to represent the three-phase equilibria. The solubility limits for the $\beta$ phase were taken from desorption curves.

Plausible initial values for the model parameters were chosen based on experience and physico-chemical intuition and evaluated while giving selected pieces of data certain weights in order to achieve satisfactory description of the system. In general, higher weights were attributed to equilibria involving three phases, and lower weights to data containing some discrepancies and inconsistencies.

%Reasonable initial values for the model parameters were chosen and evaluated according to %physical reasonableness. From a personal judgment, certain weights were given to different %pieces of selected data: higher weights to equilibria involving three phases and lower %weights when there were discrepancies and inconsistencies until a good description was %obtained. 

\section{Thermodynamic modeling}

\subsection{Para-equilibrium description}\label{thermomodelpara-eq}

To account for the para-equilibrium, we merged Fe and Ti into a hypothetical element 'FT' and set up the utilized models accordingly. The section FeTi-H of the ternary system correlates to the pseudo-binary metal-hydrogen system FT-H. This description permits the evaluation of hydride parameters under para-equilibrium since it avoids including the more stable binary phases that compose the complete equilibrium. I.e., instead of predicting the decomposition of the hydrides into Fe$_{2}$Ti + TiH$_{2}$, the equilibrium calculation describes the metastability of the ternary hydrides with the gas.

\subsection{Calculation of formation energies}
As mentioned before, the crystal structure of the $\gamma$ phase is still under debate. Therefore, the calculated lowest-energy dihydride structure predicted by ``Materials Project''developed by Jain et al. \cite{Jain2013} was employed in this work to represent $\gamma$. Based on the $\alpha$, $\beta$, and $\gamma$ structures, new compounds were produced by permuting hydrogen and vacancy in the sites that are known to allow solubility.

As an initial comparison, the DFT ground-state formation energy ($\Delta E^{\phi}_{\mathrm{FeTiH_{n}}}$) for each compound was calculated in relation to the stable-element reference (SER) as shown in Eq.~\ref{eq:groundstatephases}:

\begin{dmath}\label{eq:groundstatephases}
\Delta E^{\phi}_{\mathrm{FeTiH}_{n}} = E^{\phi}_{\mathrm{FeTiH}_{n}} - (E_{\mathrm{Fe}}^{\mathrm{bcc}} + E_{\mathrm{Ti}}^{\mathrm{hcp}} + \frac{n}{2}E_{\mathrm{H_{2}}}^{\mathrm{0K}}) \, .
\end{dmath}
  
Hydrogen gas is approximated via the DFT energy of an H$_{2}$ molecule in vacuum at 0~K (see method section), neglecting its entropic term.

\subsection{The gas phase}

The gas phase was considered to have an ideal behavior within the temperature and pressure range of interest \cite{Joubert2010}. In principle, the ternary gas can contain Fe, FeH, Fe$_{2}$, H, H$_{2}$, Ti and Ti$_{2}$. In our study, only  FT and H$_{2}$ were chosen to constitute the gas phase because these species are the most likely to be encountered in the temperature regime of hydrogen storage materials. Since H$_{2}$ is much more stable in H-rich regions and because allowing a solubility of FT supports numerical convergence during the equilibrium calculations, we adopted FT as a species in the gas phase. The total Gibbs energy of one mole of ``formula unit'' (FT, H$_{2}$) of the gas phase is then given by:

\begin{dmath}\label{eq:1}
G^{\mathrm{gas}} = \sum_{i} y_{i}[{}^{_{0}}G^{\mathrm{gas}}_{i} + RT \ln(P/P_{0})]+RT\sum_{i} y_{i} \ln y_{i} \, ,        
\end{dmath}

where ${}^{0}G^{\mathrm{gas}}_{i}$ is the standard Gibbs energy of constituent \emph{i} in the gas state from the SGTE database \cite{Dinsdale1991}, which is based on the JANAF thermochemical tables \cite{NationalInstituteofStandardsandTechnology1998}, $P_{0}$ stands for the atmospheric pressure of 101325~Pa, $y_{i}$ is the constituent fraction of species \emph{i}, and $R$ is the universal gas constant. The vapor Fe parameters were chosen to account for the hypothetical FT species in the gas phase. 

\subsection{The $\alpha$ phase} \label{alpha}

Since the B2 structure has three octahedral sites per metallic atom, a two-sublattice (2SL) model (FT)(vac, H)$_{3}$ was used to describe the $\alpha$ phase accounting for a 1:3 ratio of substitutional to interstitial sites keeping consistency with most models describing bcc-like structures \cite{Lukas2007}. The first SL accounts for metallic sites, while the second SL represents the octahedral sites where vacancies (vac) and hydrogen (H) mix. The Gibbs energy per mole of formula unit is expressed as follows:

\begin{dmath}\label{eq:2}
 G^{\alpha} = {}^{_{o}}y_{_{\mathrm{vac}}}{}^{_{0}}G^{\alpha}_{\mathrm{FT:vac}} + {}^{_{o}}y_{_{\mathrm{H}}}{}^{_{0}}G^{\alpha}_{\mathrm{FT:H}} + 3RT({}^{_{o}}y_{_{\mathrm{H}}}\ln {}^{_{o}}y_{_{\mathrm{H}}}+{}^{_{o}}y_{_{\mathrm{vac}}}\ln {}^{_{o}}y_{_{\mathrm{vac}}}) + {}^{_{o}}y_{_{\mathrm{vac}}}{}^{_{o}}y_{_{\mathrm{H}}}\sum_{\upsilon=0}^{n} {}^{\upsilon}L^{\alpha}_{\mathrm{FT:H,vac}}({}^{_{o}}y_{_{\mathrm{H}}}-{}^{_{o}}y_{_{\mathrm{vac}}})^{\upsilon} \, .
\end{dmath}

The ${}^{o}y_{_{i}}$ variables represent the fraction of species \emph{i} placed in the octahedral sites. The last term of Eq.~\ref{eq:2} is the Redlich-Kister polynomial \cite{Redlich1948}, where ${}^{\upsilon}L_{\mathrm{FT:H,vac}}$ accounts for the $\upsilon^{th}$-order binary interaction between hydrogen and vacancy in the interstitial lattice. This can be expressed as ${}^{\upsilon}L^{\alpha}_{\mathrm{FT:H,vac}} = {}^{\upsilon}A^{\alpha} + {}^{\upsilon}B^{\alpha}T$, where \emph{A} and \emph{B} are optimized model parameters (see Tab.~\ref{Table:thermoparameters}).  

The ${}^{_{0}}G^{\alpha}_{\mathrm{FT:vac}}$ and ${}^{_{0}}G^{\alpha}_{\mathrm{FT:H}}$ terms are the Gibbs energies of the end-members of the $\alpha$ phase.

${}^{_{0}}G^{\alpha}_{\mathrm{FT:vac}}$ represents the Gibbs energy of the FeTi-B2 phase. The value was obtained by analytically solving the bcc order-disorder model from Santhy and Kumar \cite{Santhy2021} assuming perfect ordering. Because the ordering is a second-order transition, the order-disorder description of the bcc phase enforces the treatment of the ordered and disordered state within the same Gibbs-energy expression. In this case, a 2SL model (Fe,Ti)(vac)$_{3}$ representing a solid solution of Fe and Ti in the bcc metallic lattice (disordered phase) and a 3SL model (Fe,Ti)(Fe,Ti)(vac)$_{3}$ to represent the partitioning of the metallic site into two different sites (ordered phase) are used. 

The disordered state of the ordered phase is given when each element fraction in each partitioned sublattice is equal to the overall mole fraction. The reduction of the Gibbs-energy due to ordering is calculated by the difference between the properly ordered phase and the `ordered phase as disordered'. The reader is referred to \cite{Lukas2007} for a more detailed explanation.

Using this model, the maximum ordering contribution in the FeTi bcc phase occurs when each partitioned metallic sublattice is fully occupied by unlike metallic atoms (see Fig.~\ref{fig:Structures} a), i.e. when the thermodynamic model possesses (Fe)$_{0.5}$(Ti)$_{0.5}$(vac)$_{3}$ or (Ti)$_{0.5}$(Fe)$_{0.5}$(vac)$_{3}$ configuration and equimolar quantities:

\begin{dmath}\label{eq:3}
{}^{_{0}}G^{\alpha}_{\mathrm{FT:vac}} = 0.5G^{\mathrm{SER}}_{\mathrm{Fe}} + 0.5G^{\mathrm{bcc}}_{\mathrm{Ti}} + 0.25{}^{_{0}}G^{\mathrm{FeTi}} 
+ 0.25G^{\mathrm{xs}}_{\mathrm{dis}} - 0.25G^{\mathrm{xs}}_{\mathrm{ord}} \, .
\end{dmath}

The parameter representing the perfect ordered bcc phase includes the formation energy of the FeTi-B2 phase referenced in the pure metals in their bcc form, and terms coming from the excess contributions of the equimolar disordered phase and the `ordered phase as disordered' (see Eq.~\ref{eq:3} and Tab.~\ref{Table:thermoparameters}). 

${}^{0}G^{\alpha}_{\mathrm{FT:H}}$ is the Gibbs energy of formation of the hypothetical compound FeTiH$_{6}$ with metallic B2 structure and all available octahedral sites occupied by H ($\alpha_{6}$). It can be expressed as:

\begin{dmath}\label{eq:4}
{}^{0}G^{\alpha}_{\mathrm{FT:H}} = {}^{_{0}}G^{\alpha}_{\mathrm{FT:vac}} + \frac{3}{2}G^{\mathrm{SER}}_{\mathrm{H_{2}}} + \Delta H_{f}^{\alpha_{6}} + \Delta S_{f}^{\alpha_{6}}T \, .
\end{dmath}

The enthalpy and entropy of formation for the FeTiH$_{6}$ compound ($\Delta H_{f}^{\alpha_{6}}$ and $\Delta S_{f}^{\alpha_{6}}$, respectively) were evaluated using the results in the present work (see Tab.~\ref{Table:thermoparameters}). The formation energy of FeTiH$_{6}$ in the $\alpha$ structure was obtained through Eq.~\ref{eq:groundstatephases} and used to represent the experimental value of $\Delta H_{f}^{\alpha_{6}}$.

%\begin{dmath}\label{eq:5}
%\Delta E^{\alpha}_{\mathrm{FeTiH_{6}}}=E^{\alpha}_{\mathrm{FeTiH_{6}}} - (E_{\mathrm{Fe}}^{\mathrm{bcc}} + E_{\mathrm{Ti}}^{\mathrm{hcp}} + 3E_{\mathrm{H_{2}}}^{GS})
%\end{dmath}

%The calculated formation energy at 0~K obtained through Eq.~\ref{eq:5} was assumed to be a reasonable approximation to the enthalpy of formation at room temperature (298.15~K) and therefore employed during the optimization of the model parameters.

\subsection{The $\beta$ phase}

Experimental evidence shows that the $\beta$ phase has at least one hydrogen per formula unit and that hydrogen occupies the first octahedral site, which is why we used a 3SL model (FT)$_{2}$(H)(vac, H) for the phase description.

The first SL represents the substitutional positions of the metallic sites, while the second and third SLs represent the H1 and H2 interstitial sites, respectively. Such a model allows additional hydrogen to enter and mix with vacancies only in H2. When the H2 sites are empty or occupied by hydrogen, the model gives rise to two end-members in the frame of $\beta$,  FeTiH and FeTiH$_{2}$, related to the $\beta_{1}$ and $\beta_{2}$ structures, respectively.

The mixing in the H2-SL allows the description of the miscibility gap between $\beta_{1}$ and $\beta_{2}$ and its dependence on temperature permits the reproduction of the observed critical temperature above which the transformation appears as a continuous process.

The Gibbs energy of $\beta$ per mole of formula unit is given as:

\begin{dmath}\label{eq:6}
 G^{\beta} = {}^{_{\mathrm{H2}}}y_{_{\mathrm{vac}}} {}^{_{0}}G^{\beta}_{\mathrm{FT:H:vac}} + {}^{_{\mathrm{H2}}}y_{_{\mathrm{H}}}{}^{_{0}}G^{\beta}_{\mathrm{FT:H:H}}
 + RT({}^{_{\mathrm{H2}}}y_{_{\mathrm{H}}}\ln {}^{_{\mathrm{H2}}}y_{_{\mathrm{H}}}+{}^{_{\mathrm{H2}}}y_{_{\mathrm{vac}}}\ln {}^{_{\mathrm{H2}}}y_{_{\mathrm{vac}}}) + {}^{_{\mathrm{H2}}}y_{_{\mathrm{vac}}}{}^{_{\mathrm{H2}}}y_{_{\mathrm{H}}}\sum_{\upsilon=0}^{n} {}^{\upsilon}L^{\beta}_{\mathrm{FT:H:H,vac}}({}^{_{\mathrm{H2}}}y_{_{\mathrm{H}}}-{}^{_{\mathrm{H2}}}y_{_{\mathrm{vac}}})^{\upsilon} \,,
\end{dmath}

where ${}^{\mathrm{_{H2}}}y_{_{i}}$ is the fraction of component \emph{i} in the H2 octahedral sites.

Analogous to the $\alpha$ phase, the last term of Eq.~\ref{eq:6} is the \emph{$L^{\beta}$} Redlich-Kister polynomial with ${}^{\upsilon}L^{\beta}_{\mathrm{FT:H:H,vac}} = {}^{\upsilon}$A${}^{\beta} + {}^{\upsilon}$B${}^{\beta}$T.

%are optimized in this work.

The ${}^{_{0}}G^{\beta}_{\mathrm{FT:H:vac}}$ and ${}^{_{0}}G^{\beta}_{\mathrm{FT:H:H}}$ terms are, respectively, the Gibbs energies of formation of $\beta_{1}$-FeTiH and  $\beta_{2}$-FeTiH$_{2}$ compounds referenced to the FeTi-B2 and H$_{2}$-gas phases:

\begin{dmath}\label{eq:7}
{}^{_{0}}G^{\beta}_{\mathrm{FT:H:vac}} = 2{}^{_{0}}G^{\alpha}_{\mathrm{FT:vac}} + \frac{1}{2}G^{\mathrm{SER}}_{\mathrm{H_{2}}} + \Delta H_{f}^{\beta_{1}} + \Delta S_{f}^{\beta_{1}}T
\end{dmath}

\begin{dmath}\label{eq:8}
{}^{_{0}}G^{\beta}_{\mathrm{FT:H:H}} = 2{}^{_{0}}G^{\alpha}_{\mathrm{FT:vac}} + G^{\mathrm{SER}}_{\mathrm{H_{2}}} + \Delta H_{f}^{\beta_{2}} + \Delta S_{f}^{\beta_{2}}T \,.
\end{dmath}

$\Delta H_{f}^{*}$ and $\Delta S_{f}^{*}$ are evaluated parameters (see Tab.~\ref{Table:thermoparameters}). %parameters to be optimized.

For the end-members of the $\beta$ phase, the DFT formation energies were calculated in reference to the B2 phase and the H$_{2}$ molecule, as stated in Eqs.~\ref{eq:9} and \ref{eq:10}:

%The end-member energies were calculated via first principals and the formation energy of the $\beta_{1}$ and $\beta_{2}$ phases were derived by the difference of the total energy of the compounds from FeTi-B2 and H$_{2}$ molecule at 0~K:

\begin{dmath}\label{eq:9}
\Delta E^{\beta_{1}}_{\mathrm{FeTiH}}= E^{\beta_{1}}_{\mathrm{FeTiH}} - (E_{\mathrm{FeTi}}^{\mathrm{B2}} + \frac{1}{2}E_{\mathrm{H_{2}}}^{0K})
\end{dmath}

\begin{dmath}\label{eq:10}
\Delta E^{\beta_{2}}_{\mathrm{FeTiH_{2}}}= E^{\beta_{2}}_{\mathrm{FeTiH_{2}}} - (E_{\mathrm{FeTi}}^{\mathrm{B2}} + E_{\mathrm{H_{2}}}^{\mathrm{0K}}) \,.
\end{dmath}

%Again, the calculated energies from Eq.~\ref{eq:9} and Eq.~\ref{eq:10} were treated as enthalpies of formation at 298.15~K during the assessment of model parameters.

%\subsection{The $\gamma$ phase}

%The first-principals analysis of the formation energy of the $\gamma$ phase was conducted similarly as previously stated:

%\begin{dmath}\label{eq:11}
%\Delta E^{\gamma}_{FeTiH_{2}}= E^{\gamma}_{FeTiH_{2}} - (E_{FeTi}^{B2} + E_{H_{2}}^{0K})
%\end{dmath}

%A thermodynamic model for the $\gamma$ phase was disregarded. The reasons for this are explained in the next section.

\section{Results and Discussion}\label{results}

\subsection{Ground state analysis}

\begin{figure}[h!] 
    \centering
    \includegraphics[width=8cm]{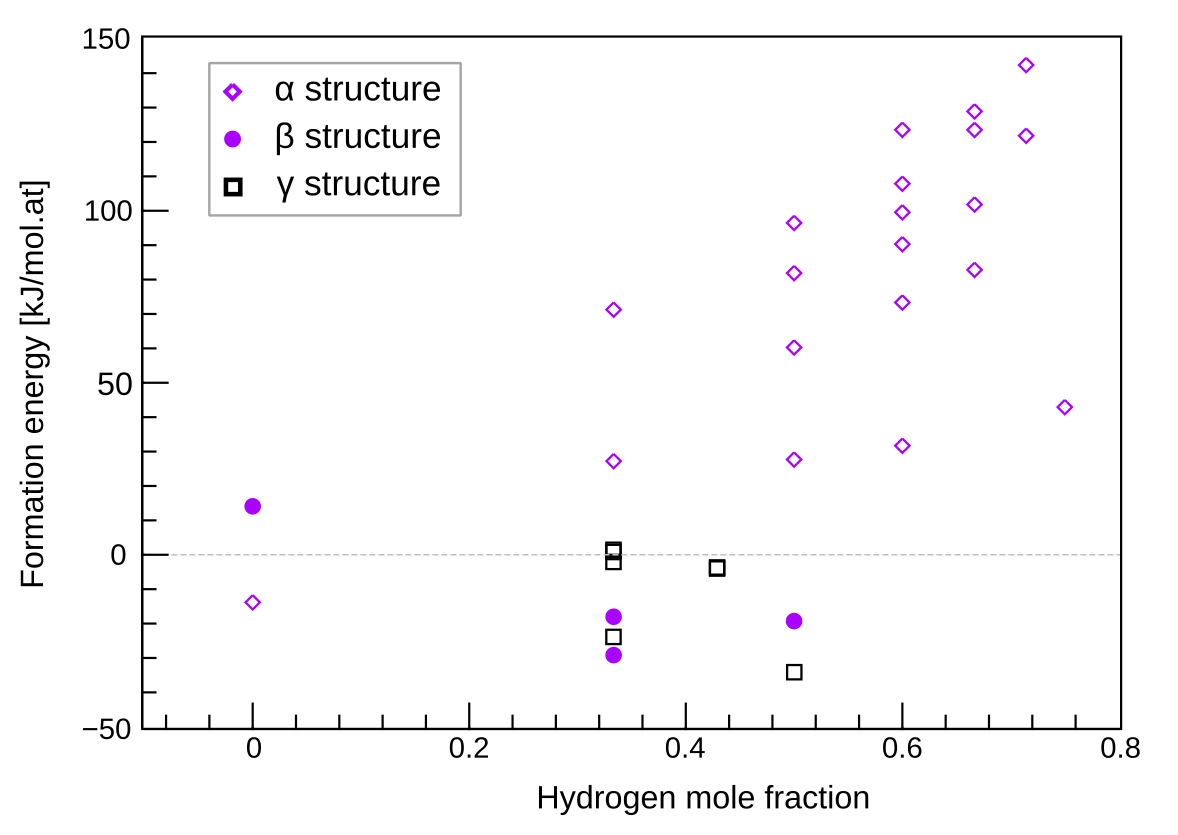}
    \caption{DFT ground state formation energies of $\alpha$, $\beta$ and $\gamma$ structures obtained via hydrogen-vacancy permutations in the octahedral sites. The large spread of values for $\alpha$ result from the combination of low hydrogen solubility in this phase and the large number of possible occupation sites for H in this structure.}
    \label{fig:form-energies}
\end{figure}

The ground state formation energies of the analyzed compounds are summarized in Tab.~\ref{Table:structure&energies} and shown in Fig.~\ref{fig:form-energies}.

\begin{table}[htp]
\caption{Compound models, DFT ground state formation energies ($\Delta$E$_{f}$) and calculated enthalpies of formation ($\Delta$H$_{f}$) per mole of atom at 298~K.}
%\resizebox{\textwidth}{!}{%
\centering
\begin{adjustbox}{width=7.8cm}
\begin{tabular}{>{\columncolor[gray]{0.92}}p{5em}>{\columncolor[gray]{0.92}}p{3em}>{\columncolor[gray]{0.92}}p{7em}>{\columncolor[gray]{0.92}}p{3em}>{\columncolor[gray]{0.92}}p{4em}} %change to what you like
\hline
%\space & \space & \space & \space & \space \\
Compound & Space group & Thermodynamic model &  $\Delta E_{f}$ [kJ/mol] & $\Delta H_{f}$ [kJ/mol] \\
%\space & \space & \space & \space & \space \\
\hline
$\alpha-$FeTiH$_{6}$ & Pm3m & (FT)(H,vac)$_{3}$ & 43.8 & 43.7 \\

%$\beta_{1}-$FeTiH & Pmma & \multirow{2}{*}{(FT)$_{2}$(H)(H,vac)}   & -20.0 & -20.1 \\
$\beta_{1}-$FeTiH & &  & --20.0 & --20.1 \\

$\beta_{2}-$FeTiH$_{2}$ &\multirow{-2}{*}{\cellcolor[gray]{.92}Pmma}  & \multirow{-2}{*}{\cellcolor[gray]{.92}(FT)$_{2}$(H)(H,vac)} & --12.4 & --12.5 \\
%$\beta_{2}-$FeTiH$_{2}$ & Pmma &  & -12.4 & -12.5 \\
$\gamma-$FeTiH$_{2}$ & Cmmm & --- & --27.3 & --- \\
\hline
\end{tabular}%}
\end{adjustbox}
\label{Table:structure&energies}
\end{table}

Even though the mole fraction solubility limit of hydrogen in $\alpha$ is around 0.047, H may occupy any of the available six octahedral sites.
However, note that the addition of a single H atom into the  $\alpha$ unit cell already surpasses the experimentally observed solubility limit. Consequently, the generated $\alpha$ structures containing hydrogen correspond to hypothetical and unstable compounds with positive energies of formation.

Due to symmetry, there are only two options for adding a single hydrogen to the $\alpha$-phase unit cell: the Fe$_{2}$Ti$_{4}$ and the Fe$_{4}$Ti$_{2}$ octahedral sites. The permutation of distributing up to six H atoms in both octahedral sites produces 20 distinguished compounds. Our calculations (see Table S1 in the Supporting Information) show that the lowest formation energy for adding a given number of H atoms into the $\alpha$ unit cell lattice is obtained when the H atoms are the furthest apart from each other as possible and preferentially located in Fe$_{2}$Ti$_{4}$ sites, agreeing with the findings from Nong et al. \cite{Nong2014}.

The construction of $\beta$ structures involved the permutation of two H atoms and vacancies in the H1 and H2 Fe$_{2}$Ti$_{4}$ interstitial sites, which are known to permit some H solubility. The $\beta$-FeTiH structure with the lowest energy agrees to that provided by ``Materials Project'' \cite{Jain2013}. Interestingly, all the analyzed $\beta$-structures containing H have lower molar energy of formation at 0~K than the pure $\alpha$-FeTi.

As mentioned in section \ref{hydrogenationsec}, there is some experimental evidence regarding competing stability between $\gamma$ and $\beta_{2}$. Our \emph{ab-initio} results suggest $\gamma$ to possess highest thermodynamic stability as it has the lowest calculated molar formation energy among all analyzed hydrides in the FeTi-H system. 

Moreover, our DFT calculations reveal that upon removal of one H per formula unit from the $\gamma$ unit cell, including those located in Fe$_{4}$Ti$_{2}$ octahedral sites, the structure tends to relax, recovers orthogonality and approaches cell shape and formation energy of $\beta_{1}$.

%Due to its higher stability, including the $\gamma$ phase formation energy into the model either with stoichiometric description or with a 3SL model would require unrealistic entropy of formation to reproduce the experimental data. Therefore, we chose not to include the $\gamma$ phase model.

On the other hand it is interesting to note that the $\beta$ dihydride formation energy is also negative, suggesting that a dihydride with $\beta$ structure may form as a metastable state in the system. By using the calculated $\beta_{2}$ formation energy to represent the end-member of $\beta$, we were able to reproduce the experimental PCI data more closely. Thus, it appears to be a more realistic description and we hypothesize that the formation of $\beta$ during the hydrogenation might be a metastable process inside the frame of para-equilibrium. For this reason, we chose not to include a $\gamma$ phase model in the thermodynamic description at this point as our approach already reproduced the underlying physico-chemical processes relevant for technical application of FeTi as a storage material faithfully with respect to experimental observations. However, a more detailed description including a $\gamma$ phase model will follow in the near future.

We can further assume that upon hydrogenation, the formation of $\beta$ facilitates the hydrogen atoms to first diffuse into the Fe$_{2}$Ti$_{4}$ sites, forming $\beta_{2}$. Subsequently, if sufficient external conditions are given to surpass the activation energy for the hydrogen to jump into Fe$_{4}$Ti$_{2}$ sites, $\beta_{2}$ transforms into $\gamma$.
This mechanism supports the explanation of why $\beta_{2}$ is observed during absorption, in contrast to $\beta_{1}$ that appears only during desorption \cite{Reidinger1982} as well as the high hysteresis level upon absorption-desorption cycles \cite{Goodell1980}.

\subsection{Thermodynamic evaluation}

The auxiliary functions and assessed parameters optimized from the experimental data are listed in Tab.~\ref{Table:thermoparameters}.

\begin{table}[htp]
\caption{Assessed thermodynamic parameters of the FeTi-H system.}
%\resizebox{\textwidth}{!}{%
\centering
\begin{adjustbox}{width=7.8cm}
\begin{tabular}{>{\columncolor[gray]{0.92}}p{18em}>{\columncolor[gray]{0.92}}p{8em}} %change to what you like
\hline
\space & \space \\
\textbf{Model parameters} & \textbf{Reference} \\
\space & \space \\
\hline
\space & \space \\
$G^{\mathrm{xs}}_{\mathrm{dis}}= -68448 +23.825T $ & \cite{Santhy2021} and this work  \\
$G^{\mathrm{xs}}_{\mathrm{ord}}= -10953 -6097 $ & \cite{Santhy2021} and this work  \\
${}^{_{0}}G^{\mathrm{FeTi}}= -76147-46.603T+8.663T\ln(T)-7.151E^{-3}T^{2} +1.121169E^{-6}T^{3} $ & \cite{Santhy2021}  \\
${}^{_{0}}G^{\mathrm{B2}}_{\mathrm{Fe:Ti:vac}}=0.5{}^{_{0}}G^{\mathrm{FeTi}} $ & \cite{Santhy2021} and this work  \\[1ex]
${}^{_{0}}G^{\alpha}_{\mathrm{FT:vac}} = 0.5G^{\mathrm{SER}}_{\mathrm{Fe}} + 0.5G^{\mathrm{bcc}}_{\mathrm{Ti}} + 0.25{}^{_{0}}G^{\mathrm{FeTi}} 
+ 0.25G^{\mathrm{xs}}_{\mathrm{dis}} - 0.25G^{\mathrm{xs}}_{\mathrm{ord}}$ & \cite{Santhy2021} and this work  \\[1ex]
${}^{_{0}}G^{\alpha}_{\mathrm{FT:H}} = {}^{_{0}}G^{\alpha}_{\mathrm{FT:vac}} + \frac{3}{2}G^{\mathrm{SER}}_{\mathrm{H_{2}}} + 200000 
+ 130T$ & This work \\[1ex]
${}^{_{0}}G^{\beta}_{\mathrm{FT:H:vac}} = 2~{}^{_{0}}G^{\alpha}_{\mathrm{FT:vac}} + \frac{1}{2}G^{\mathrm{SER}}_{\mathrm{H_{2}}}-11160+46.7T$ & This work \\[1ex]
${}^{_{0}}G^{\beta}_{\mathrm{FT:H:H}} = 2~{}^{_{0}}G^{\alpha}_{\mathrm{FT:vac}} + G^{\mathrm{SER}}_{\mathrm{H_{2}}} -721 +35.8T$  & This work   \\[1ex]
${}^{0}L^{\alpha}_{\mathrm{FT:H,vac}} = -346135 $   & This work   \\
${}^{1}L^{\alpha}_{\mathrm{FT:H,vac}} = -149824 $   & This work    \\
${}^{0}L^{\beta}_{\mathrm{FT:H,vac}} = -20563 + 63.3T$  & This work  \\
${}^{1}L^{\beta}_{\mathrm{FT:H,vac}} = -35321 + 95T$  & This work     \\
${}^{2}L^{\beta}_{\mathrm{FT:H,vac}} = -27123 + 68.7T$   & This work    \\
\space & \space \\
\hline
\end{tabular}%}
\end{adjustbox}
\label{Table:thermoparameters}
\end{table}

\begin{figure}[htb!]    
    \centering
    \includegraphics[width=8cm]{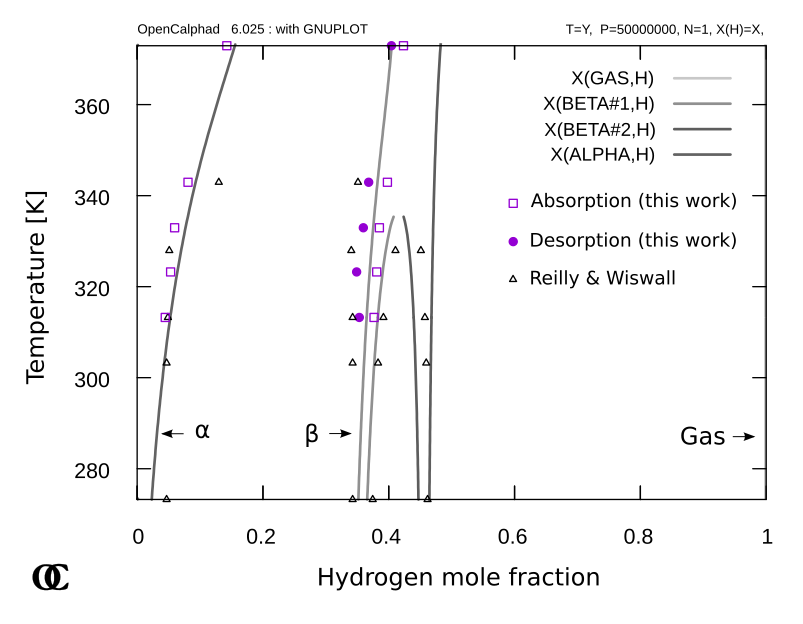}
    \caption{Calculated FeTi-H phase diagram at 50~MPa. Experimental data from this work as well as from \cite{Reilly1974} is shown for comparison.}
    \label{fig:tempcomp}
\end{figure} 

The calculated phase diagram of the FeTi-H system comparing the experimental data from this work and Reilly \cite{Reilly1974} is shown in Fig.~\ref{fig:tempcomp} . The data describes the solubility limits and the miscibility gap within the $\beta$ phase region. The estimated critical point is calculated at  $x_{_{\mathrm{H}}} =$~0.415, $T_{c} =$~336~K in good agreement with estimates by Reilly and Wiswall \cite{Reilly1974}. 

\begin{figure}[htb!]   
    \centering
    \includegraphics[width=8cm]{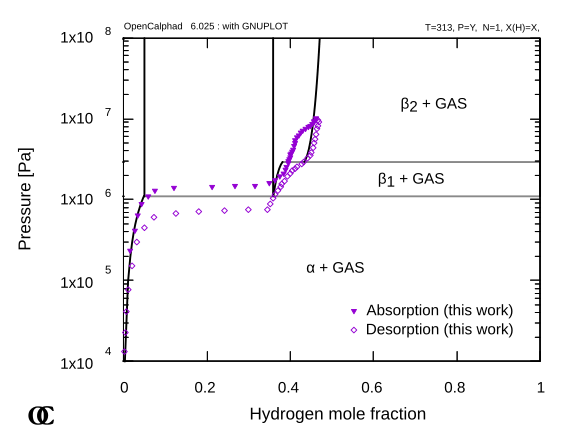}
    \caption{Calculated pressure-composition diagram in the FeTi-H system at 313~K. Superimposed is correspondent experimental PCI data from this work.}
    \label{fig:pressurecomp}
\end{figure}

The calculated pressure-composition phase diagram at 313~K superimposed with experimental data from this work is shown in Fig.~\ref{fig:pressurecomp}. Good agreement is also obtained in the analyzed pressure range, especially for the $\alpha$ solubility limits.

\begin{figure}[htb!]  
    \centering
    \includegraphics[width=8cm]{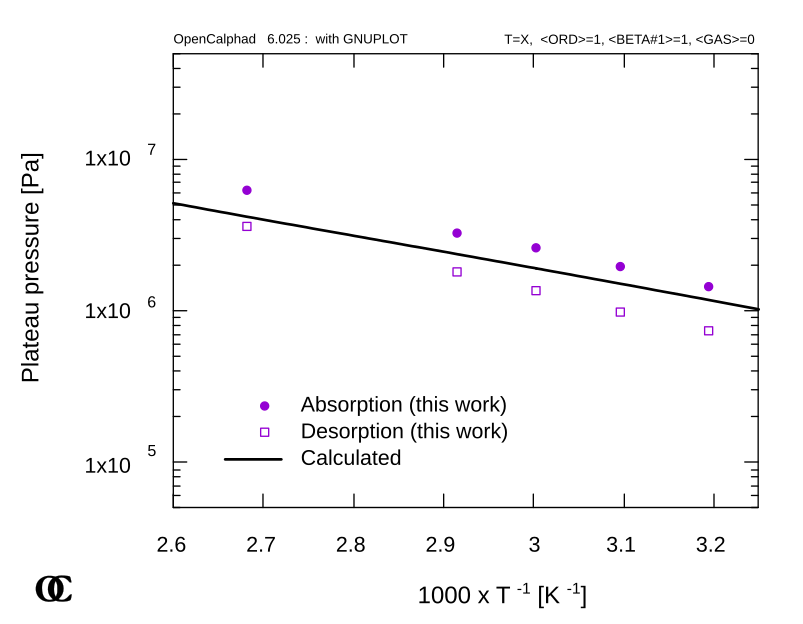}
    \caption{Calculated first plateau pressure as a function of the inverse temperature in the FeTi-H system in comparison with experimental data from this work. As to be expected, the true equilibrium plateau pressure lies between absorption and desorption curves \cite{Flanagan2005}.}
    \label{fig:plateaupressure}
\end{figure}

\begin{figure}[htb!]  
    \centering
    \includegraphics[width=8cm]{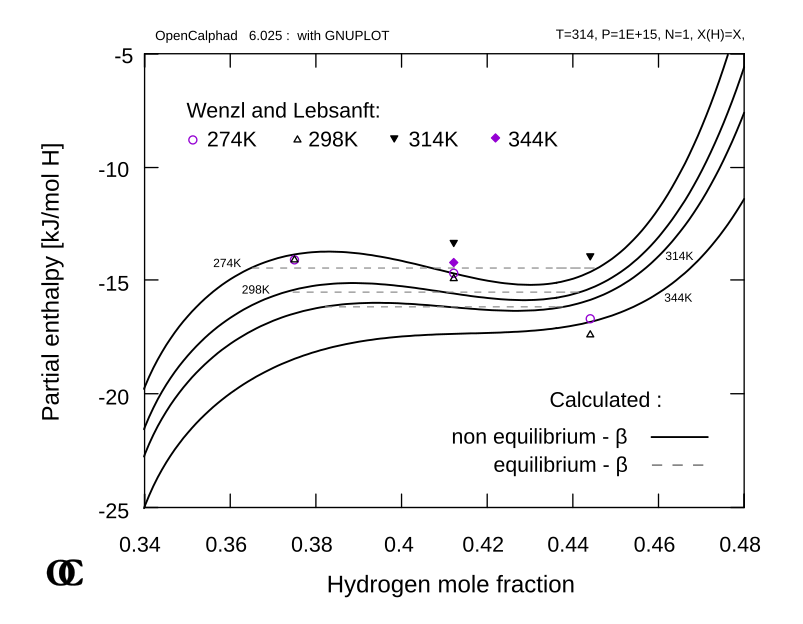}
    \caption{Calculated partial enthalpy at low temperatures as a function of composition in the FeTi–H system. Experimental data from \cite{Wenzl1980} is shown for comparison.}
    \label{fig:partialenthalpy}
\end{figure}

In Fig.~\ref{fig:plateaupressure}, the calculated equilibrium pressure is plotted together with values obtained during absorption and desorption, as it is known that the true equilibrium plateau pressure may lie between absorption and desorption curves \cite{Flanagan2005}. Fig.~\ref{fig:partialenthalpy} shows calculated partial enthalpies as a function of hydrogen fraction at temperatures below and above the miscibility gap compared to experimental data. %During assessment of the model parameters we found that using desorption curves for the solubility limits of $\beta$ enhanced equilibrium convergence in the Calphad calculations. This finding further supports the hypothesis that $\beta_{2}$ formation is facilitated during absorption.

%The calculated heat capacity as function of composition compared to the data from Wenzl and Pietz \cite{Wenzl1980a} is also shown in Fig.~\ref{fig:cp}.

%\begin{figure}[h!]  
%    \centering
%    \includegraphics[width=8cm]{Images/CP_eq_ink.png}
%    \caption{Calculated heat capacity at 300~K as a function of %composition in the FeTi–H system. Experimental data from %\cite{Wenzl1980a} is shown for comparison.}
%    \label{fig:cp}
%\end{figure} 

Although the plotted data of partial enthalpies were not taken into account during optimization, the calculated values lie in the expected range. From Wenzl and Pietz \cite{Wenzl1980a}, the heat capacity at constant pressure ($C_P$) of Fe$_{0.49}$Mn$_{0.01}$Ti$_{0.5}$H$_{0.47}$ (corresponding to  $\beta_{1}$ composition), is measured to be 18.3~J/mol$\cdot$K at 280~K and may be compared to our calculated $C_P$ value of 20.5~J/mol$\cdot$K of the $\beta_{1}$-FeTiH phase at the same temperature. Since phase transformations occurring at high hydrogenation level (surpassing $\beta_{1}$ composition) may not reach the complete equilibrium, data in this composition range is scarce and may depend heavily on the history of the sample (phase constitution). For this reason, we conclude by stating that our calculations agree well with the available experimental data and show that the developed models can describe the most relevant thermodynamic aspects for design and optimization of intermetallic materials for hydrogen storage.

\section{Conclusions}

We successfully assessed the thermodynamics of the FeTi-H system by properly selecting phase models and optimizing associated parameters based on a combination of
\begin{enumerate}
    \item Experimental data on the solubility limits acquired from new PCI measurements \vspace{0.2cm}
    \item DFT calculations of ground state formation energies.
\end{enumerate}

The thermodynamic description of the FeTi-H system was represented by modeling the Gibbs energies of individual phases.
Our thermodynamic assessment of the system reproduces most thermodynamic quantities that are of technological importance for development of metal hydride-based hydrogen storage materials, even though only the $\alpha$ and $\beta$ hydrides were considered in the model.\\
\vspace{1.0cm}

Additional key findings include:

\begin{itemize}
\item{} Our \emph{ab-initio} calculation results show that hydrides with  $\beta$ and $\gamma$ structures have negative formation energy when hydrogen atoms are permuted in the octahedral sites and thus are likely to be formed during hydrogenation.
\item{} The noticeably lower ground state energy of formation of $\gamma$ dihydride compared to $\beta$ dihydride suggests that the $\gamma$ phase may emerge under certain conditions, which in turn hinders the reversibility of the hydrogenation of FeTi. 
\end{itemize}

The present work is expected to serve as a key basis for modeling hydrides involving the FeTi system. In addition, the assessed thermodynamics can provide necessary input parameters for kinetic models and hydride phase microstructure simulations \cite{Heo2019}. Note here that we assessed the thermodynamics of the FeTi-H system employing the freely available OpenCalphad software, which to our knowledge is the first instance it was used for a complete thermodynamic assessment. 

In our subsequent work, we plan to extend these pseudo-binary models to account for full ternary systems, allowing for more detailed hydrogenation simulations for FeTi and other MHs.

\section*{Acknowledgement}

We are grateful to Dr. J.-M. Joubert (Institut de Chimie et des Matériaux Paris-Est, France), for stimulating and fruitful discussions.

\small{This work was partly performed under the auspices of the U.S. Department of Energy by Lawrence Livermore National Laboratory (LLNL) under Contract DE-AC52-07NA27344. The work by LLNL was supported by the Hydrogen Storage Materials Advanced Research Consortium (HyMARC) of the U.S. Department of Energy (DOE), Office of Energy Efficiency and Renewable Energy, Hydrogen and Fuel
Cell Technologies Office under Contract DE-AC52-07NA27344. The views and opinions of the authors expressed herein do not necessarily state or reflect those of the United States Government or any agency thereof. Neither the United States Government nor any agency thereof, nor any of their employees, makes any warranty, expressed or implied, or assumes any legal liability or responsibility for the accuracy, completeness, or usefulness of any information, apparatus, product, or process disclosed, or represents that its use would not infringe privately owned rights.}

\section*{Data availability}

The raw data required to reproduce these findings are available to download as Electronic Supplementary Information attached as part of this article. The processed data required to reproduce these findings are available to download as Electronic Supplementary Information attached as part of this article.

\bibliography{bibliography}

\begin{thebibliography}{10}
\expandafter\ifx\csname url\endcsname\relax
  \def\url#1{\texttt{#1}}\fi
\expandafter\ifx\csname urlprefix\endcsname\relax\def\urlprefix{URL }\fi
\expandafter\ifx\csname href\endcsname\relax
  \def\href#1#2{#2} \def\path#1{#1}\fi

\bibitem{Hirscher2010}
M.~Hirscher (Ed.), Handbook of {{Hydrogen Storage}}: {{New Materials}} for
  {{Future Energy Storage}}, {Wiley-VCH}, 2010.
\newblock \href {https://doi.org/10.1002/9783527629800}
  {\path{doi:10.1002/9783527629800}}.

\bibitem{Dematteis2021a}
E.~M. Dematteis, N.~Berti, F.~Cuevas, M.~Latroche, M.~Baricco, {Substitutional
  effects in TiFe for hydrogen storage: A comprehensive review}, Mater. Adv.
  2~(8) (2021) 2524--2560.
\newblock \href {http://arxiv.org/abs/2102.04766} {\path{arXiv:2102.04766}},
  \href {https://doi.org/10.1039/d1ma00101a} {\path{doi:10.1039/d1ma00101a}}.

\bibitem{Lys2020}
A.~Lys, J.~O. Fadonougbo, M.~Faisal, J.-Y. Suh, Y.-S. Lee, J.-H. Shim, J.~Park,
  Y.~W. Cho, {Enhancing the Hydrogen Storage Properties of
  A\textsubscript{x}B\textsubscript{y} Intermetallic Compounds by Partial
  Substitution: A Short Review}, Hydrogen 1~(1) (2020) 38--63.
\newblock \href {https://doi.org/10.3390/hydrogen1010004}
  {\path{doi:10.3390/hydrogen1010004}}.

\bibitem{Broom2011}
D.~P. Broom, {Hydrogen Storage Materials: The Characterization of Their Storage
  Properties.}, Springer Science \& Business Media, 2011.
\newblock \href {https://doi.org/10.1007/978-0-85729-221-6}
  {\path{doi:10.1007/978-0-85729-221-6}}.

\bibitem{Sasaki2016}
K.~Sasaki, H.-W. Li, A.~Hayashi, J.~Yamabe, T.~Ogura, {Hai-Wen Li}, {Akari
  Hayashi}, {Junichiro Yamabe}, {Teppei Ogura}, {Stephen M. Lyth},
  \href{http://www.springer.com/series/8059%0Ahttp://link.springer.com/10.1007/978-4-431-56042-5}{{Hydrogen
  Energy Engineering: A Japanese Perspective}}, 2016.
\newline\urlprefix\url{http://www.springer.com/series/8059%0Ahttp://link.springer.com/10.1007/978-4-431-56042-5}

\bibitem{HyCARE2019}
{HyCARE focuses on large-scale, solid-state hydrogen storage}, Fuel Cells
  Bulletin 2 (2019) 11.
\newblock \href {https://doi.org/10.1016/s1464-2859(19)30068-9}
  {\path{doi:10.1016/s1464-2859(19)30068-9}}.

\bibitem{GKN2021}
\href{https://www.gknhydrogen.com/#pilotprojects}{{GKN Hydrogen}}.
\newline\urlprefix\url{https://www.gknhydrogen.com/#pilotprojects}

\bibitem{Kattner2016}
U.~R. Kattner, {The Calphad Method and Its Role in Material and Process
  Development}, Tecnol. Metal. Mater. Min. 13~(1) (2016) 3--15.
\newblock \href {https://doi.org/10.4322/2176-1523.1059}
  {\path{doi:10.4322/2176-1523.1059}}.

\bibitem{Lukas2007}
H.~L. Lukas, S.~G. Fries, B.~Sundman, Computational Thermodynamics: The Calphad
  Method, Cambridge University Press, 2007.
\newblock \href {https://doi.org/10.1017/CBO9780511804137}
  {\path{doi:10.1017/CBO9780511804137}}.

\bibitem{Fukai2005}
Y.~Fukai, The Metal-Hydrogen System: Basic Bulk Properties, Springer, Berlin,
  2005, pp. 9--53.
\newblock \href {https://doi.org/doi.org/10.1007/3-540-28883-X}
  {\path{doi:doi.org/10.1007/3-540-28883-X}}.

\bibitem{Zinkevich2002}
M.~Zinkevich, N.~Mattern, A.~Handstein, O.~Gutfleisch, {Thermodynamics of
  Fe-Sm, Fe-H, and H-Sm systems and its application to the
  hydrogen-disproportionation-desorption-recombination (HDDR) process for the
  system Fe\textsubscript{17}Sm\textsubscript{2}-H\textsubscript{2}}, J. Alloys
  Compd. 339~(1-2) (2002) 118--139.
\newblock \href {https://doi.org/10.1016/S0925-8388(01)01990-9}
  {\path{doi:10.1016/S0925-8388(01)01990-9}}.

\bibitem{Antonov1981}
V.~Antonov, I.~Belash, E.~Ponyatovskii, V.~Thiessen, V.~Shiryaev,
  {Magnetization of Iron Hydride}, Phys. Status Solidi A 65~(1) (1981)
  K43--K48.
\newblock \href {https://doi.org/10.1002/pssa.2210650156}
  {\path{doi:10.1002/pssa.2210650156}}.

\bibitem{Wille1981}
G.~W. Wille, J.~W. Davis, {Hydrogen in Titanium Alloys}, Tech. rep., McDonnell
  Douglas Astronautics Co., St. Louis, MO (USA) (1981).
\newblock \href {https://doi.org/10.2172/6420120} {\path{doi:10.2172/6420120}}.

\bibitem{San-Martin1987}
A.~San-Martin, F.~D. Manchester, {The H-Ti (Hydrogen-Titanium) system}, Bull.
  Alloy Phase Diagr. 8~(1) (1987) 30--42.
\newblock \href {https://doi.org/10.1007/BF02868888}
  {\path{doi:10.1007/BF02868888}}.

\bibitem{Arita1982}
M.~Arita, K.~Shimizu, Y.~Ichinose, {Thermodynamics of the Ti-H system}, Metall.
  Trans. A 13~(8) (1982) 1329--1336.
\newblock \href {https://doi.org/10.1007/BF02642869}
  {\path{doi:10.1007/BF02642869}}.

\bibitem{Thompson1980}
P.~Thompson, F.~Reidinger, J.~J. Reilly, L.~M. Corliss, J.~M. Hastings,
  {Neutron diffraction study of $\alpha$-iron titanium deuteride}, J. Phys. F:
  Met. Phys. 10~(2) (1980) L57--L59.
\newblock \href {https://doi.org/10.1088/0305-4608/10/2/001}
  {\path{doi:10.1088/0305-4608/10/2/001}}.

\bibitem{DaSilva1976}
J.~R. da~Silva, S.~W. Stafford, R.~B. McLellan, {The thermodynamics of the
  hydrogen-iron system}, J. Less-common Metals 49 (1976) 407--420.
\newblock \href {https://doi.org/10.1016/0022-5088(76)90052-7}
  {\path{doi:10.1016/0022-5088(76)90052-7}}.

\bibitem{Jiang2004}
D.~E. Jiang, E.~A. Carter, {Diffusion of interstitial hydrogen into and through
  bcc Fe from first principles}, Phys. Rev. B 70~(6) (2004) 064102.
\newblock \href {https://doi.org/10.1103/PhysRevB.70.064102}
  {\path{doi:10.1103/PhysRevB.70.064102}}.

\bibitem{Beck1975}
F.~H. Beck, \href{https://ntrs.nasa.gov/citations/19750018024}{{Effect of
  Hydrogen on the Mechanical Properties of titanium and its Alloys}}, {NASA
  Contractor Report NASA-CR-134796} (1975).
\newline\urlprefix\url{https://ntrs.nasa.gov/citations/19750018024}

\bibitem{Reilly1974}
J.~J. Reilly, R.~H. Wiswall, {Formation and properties of iron titanium
  hydride}, Inorg. Chem. 13~(1) (1974) 218--222.
\newblock \href {https://doi.org/10.1021/ic50131a042}
  {\path{doi:10.1021/ic50131a042}}.

\bibitem{Thompson1978}
P.~Thompson, M.~A. Pick, F.~Reidinger, L.~M. Corliss, J.~M. Hastings, J.~J.
  Reilly, {Neutron diffraction study of $\beta$ iron titanium deuteride}, J.
  Phys. F: Met. Phys. 8~(4) (1978) L75--L80.
\newblock \href {https://doi.org/10.1088/0305-4608/8/4/001}
  {\path{doi:10.1088/0305-4608/8/4/001}}.

\bibitem{Thompson1979}
P.~Thompson, J.~J. Reilly, F.~Reidinger, J.~M. Hastings, L.~M. Corliss,
  {Neutron diffraction study of gamma iron titanium deuteride}, J. Phys. F:
  Met. Phys. 9~(4) (1979) L61--L66.
\newblock \href {https://doi.org/10.1088/0305-4608/9/4/001}
  {\path{doi:10.1088/0305-4608/9/4/001}}.

\bibitem{Schober1979}
T.~Schober, {The iron-titanium - hydrogen system: A transmission electron
  microscope (TEM) study}, Scr. Metall. 13~(2) (1979) 107--112.
\newblock \href {https://doi.org/10.1016/0036-9748(79)90046-2}
  {\path{doi:10.1016/0036-9748(79)90046-2}}.

\bibitem{Schafer1980}
W.~Sch{\"{a}}fer, G.~Will, T.~Schober, {Neutron and electron diffraction of the
  FeTi-D(H)-$\gamma$ phase}, Mater. Res. Bull. 15~(5) (1980) 627--634.
\newblock \href {https://doi.org/10.1016/0025-5408(80)90143-9}
  {\path{doi:10.1016/0025-5408(80)90143-9}}.

\bibitem{Fischer1978}
P.~Fischer, W.~H{\"{a}}lg, L.~Schlapbach, F.~Stucki, A.~F. Andresen, {Deuterium
  storage in FeTi. Measurement of desorption isotherms and structural studies
  by means of neutron diffraction}, Mater. Res. Bull. 13~(9) (1978) 931--946.
\newblock \href {https://doi.org/10.1016/0025-5408(78)90105-8}
  {\path{doi:10.1016/0025-5408(78)90105-8}}.

\bibitem{Schober1980}
T.~Schober, W.~Schaefer, {Transmission electron microscopy neutron diffraction
  studies of FeTi-H(D)}, J. Less-common Metals 74~(1) (1980) 23--31.
\newblock \href {https://doi.org/10.1016/0022-5088(80)90070-3}
  {\path{doi:10.1016/0022-5088(80)90070-3}}.

\bibitem{Reidinger1982}
F.~Reidinger, J.~F. Lynch, J.~J. Reilly, {An x-ray diffraction examination of
  the FeTi-H2 system}, J. Phys. F: Met. Phys. 12~(3) (1982) L49--L55.
\newblock \href {https://doi.org/10.1088/0305-4608/12/3/007}
  {\path{doi:10.1088/0305-4608/12/3/007}}.

\bibitem{Nong2014}
Z.~S. Nong, J.~C. Zhu, X.~W. Yang, Y.~Cao, Z.~H. Lai, Y.~Liu, {First-principles
  study of hydrogen storage and diffusion in B2 FeTi alloy}, Comput. Mater.
  Sci. 81 (2014) 517--523.
\newblock \href {https://doi.org/10.1016/j.commatsci.2013.08.060}
  {\path{doi:10.1016/j.commatsci.2013.08.060}}.

\bibitem{Schefer1979}
J.~Schefer, P.~Fischer, W.~H{\"{a}}lg, F.~Stucki, L.~Schlapbach, A.~F.
  Andresen, {Structural phase transitions of FeTi-deuterides}, Mater. Res.
  Bull. 14~(10) (1979) 1281--1294.
\newblock \href {https://doi.org/10.1016/0025-5408(79)90005-9}
  {\path{doi:10.1016/0025-5408(79)90005-9}}.

\bibitem{Schafer1979}
W.~Sch{\"{a}}fer, E.~Lebsanft, A.~Bl{\"{a}}sius, {Investigation of TiFe
  Deuteride Structures by Means of Neutron Powder Diffraction and the
  Mössbauer Effect}, Z. Phys. Chem. 115 (1979) 201--212.
\newblock \href {https://doi.org/10.1524/zpch.1979.115.2.201}
  {\path{doi:10.1524/zpch.1979.115.2.201}}.

\bibitem{Fischer1987}
P.~Fischer, J.~Schefer, K.~Yvon, L.~Schlapbach, T.~Riesterer, {Orthorhombic
  structure of $\gamma$-TiFeD$_{\approx2}$}, J. Less-common Metals 129 (1987)
  39--45.
\newblock \href {https://doi.org/10.1016/0022-5088(87)90031-2}
  {\path{doi:10.1016/0022-5088(87)90031-2}}.

\bibitem{Thompson1989}
P.~Thompson, J.~J. Reilly, J.~M. Hastings, {The application of the Rietveld
  method to a highly strained material with microtwins:
  TiFeD\textsubscript{1.9}}, J. Appl. Cryst. 22 (1989) 256--260.
\newblock \href {https://doi.org/10.1107/s002188988801430x}
  {\path{doi:10.1107/s002188988801430x}}.

\bibitem{Flanagan2005}
T.~B. Flanagan, W.~A. Oates, {Some thermodynamic aspects of metal hydrogen
  systems}, J. Alloys Compd. 404-406 (2005) 16--23.
\newblock \href {https://doi.org/10.1016/j.jallcom.2004.11.108}
  {\path{doi:10.1016/j.jallcom.2004.11.108}}.

\bibitem{Joubert2012}
J.~M. Joubert, {CALPHAD Modeling of Metal–Hydrogen Systems: A Review}, JOM
  64~(12) (2012) 1438--1447.
\newblock \href {https://doi.org/10.1007/s11837-012-0462-6}
  {\path{doi:10.1007/s11837-012-0462-6}}.

\bibitem{Tessier1995}
P.~Tessier,
  \href{https://escholarship.mcgill.ca/concern/theses/w0892c43c}{{Hydrogen
  Storage in Metastable Fe-Ti}}, Ph.D. thesis, McGill University (1995).
\newline\urlprefix\url{https://escholarship.mcgill.ca/concern/theses/w0892c43c}

\bibitem{Wenzl1980}
H.~Wenzl, E.~Lebsanft, {Phase diagram and thermodynamic parameters of the
  quasibinary interstitial alloy
  Fe\textsubscript{0.5}Ti\textsubscript{0.5}H\textsubscript{x} in equilibrium
  with hydrogen gas}, J. Phys. F: Met. Phys. 10~(10) (1980) 2147--2156.
\newblock \href {https://doi.org/10.1088/0305-4608/10/10/012}
  {\path{doi:10.1088/0305-4608/10/10/012}}.

\bibitem{Yamanaka1975}
K.~Yamanaka, H.~Saito, M.~Someno, {Hydride Formation of Intermetallic Compounds
  of Titanium-iron, -Cobalt, -Nickel and -Copper}, Nippon Kagaku Kaishi
  (Journal of The Chemical Society of Japan, Chemistry and Industrial
  Chemistry)~(8) (1975) 1267--1272.
\newblock \href {https://doi.org/10.1246/nikkashi.1975.1267}
  {\path{doi:10.1246/nikkashi.1975.1267}}.

\bibitem{Sujan2020}
G.~K. Sujan, Z.~Pan, H.~Li, D.~Liang, N.~Alam, {An overview on TiFe
  intermetallic for solid-state hydrogen storage: microstructure, hydrogenation
  and fabrication processes}, Crit. Rev. Solid State Mater. Sci. 45~(5) (2020)
  410--427.
\newblock \href {https://doi.org/10.1080/10408436.2019.1652143}
  {\path{doi:10.1080/10408436.2019.1652143}}.

\bibitem{Goodell1980}
P.~D. Goodell, G.~D. Sandrock, E.~L. Huston, {Kinetic and Dynamic Aspects of
  Rechargeable Metal Hydrides.}, J. Less-common Metals 73~(1) (1980) 135--142.
\newblock \href {https://doi.org/10.1016/0022-5088(80)90352-5}
  {\path{doi:10.1016/0022-5088(80)90352-5}}.

\bibitem{BOWMAN197897}
R.~Bowman, A.~Attalla, G.~Carter, Y.~Chabre, {NMR Studies of Hydrogen
  Relaxation and Diffusion in TiFeH\textsubscript{x} and
  TiFe\textsubscript{1-y}Mn\textsubscript{y}H\textsubscript{x}}, in: A.~F.
  Andresen, A.~J. Maeland (Eds.), Hydrides for Energy Storage, Pergamon, 1978,
  pp. 97--118.
\newblock \href {https://doi.org/10.1016/B978-0-08-022715-3.50015-1}
  {\path{doi:10.1016/B978-0-08-022715-3.50015-1}}.

\bibitem{Reilly1976}
J.~J. Reilly, R.~H. Wiswall,
  \href{https://www.osti.gov/biblio/7269838}{{Hydrogen storage and purification
  systems III. [Pressure-temperature composition relationships]}}, Tech. rep.,
  Brookhaven National Lab., Upton, N.Y. (USA), BNL-21322 (1976).
\newline\urlprefix\url{https://www.osti.gov/biblio/7269838}

\bibitem{Lebsanft1978}
E.~Lebsanft, \href{http://hdl.handle.net/2128/17168}{{Das System TiFeH(D) -
  Strukturen, Phasendiagramm, Thermodynamik, Diffusion des Wasserstoffs in
  TiFeH und Methoden der Pr{\"{a}}paration}}, Ph.D. thesis,
  Kernforschungsanlage Jülich (1978).
\newline\urlprefix\url{http://hdl.handle.net/2128/17168}

\bibitem{Endo2013}
N.~Endo, H.~Saitoh, A.~Machida, Y.~Katayama, {Formation of BCC TiFe hydride
  under high hydrogen pressure}, Int. J. Hydrog. Energy 38~(16) (2013)
  6726--6729.
\newblock \href {https://doi.org/10.1016/j.ijhydene.2013.03.120}
  {\path{doi:10.1016/j.ijhydene.2013.03.120}}.

\bibitem{Sieverts1907}
A.~Sieverts, {Zur Kenntnis der Okklusion und Diffusion von Gasen durch
  Metalle}, Z. Phys. Chem. 60U~(1) (1907) 129--201.
\newblock \href {https://doi.org/10.1515/zpch-1907-6009}
  {\path{doi:10.1515/zpch-1907-6009}}.

\bibitem{Kresse1993}
G.~Kresse, J.~Hafner, {\emph{Ab initio} molecular dynamics for liquid metals},
  Phys. Rev. B 47~(1) (1993) 558--561.
\newblock \href {https://doi.org/10.1103/PhysRevB.47.558}
  {\path{doi:10.1103/PhysRevB.47.558}}.

\bibitem{Kresse1996a}
G.~Kresse, J.~Furthm\"uller, Efficient iterative schemes for \emph{ab initio}
  total-energy calculations using a plane-wave basis set, Phys. Rev. B 54
  (1996) 11169--11186.
\newblock \href {https://doi.org/10.1103/PhysRevB.54.11169}
  {\path{doi:10.1103/PhysRevB.54.11169}}.

\bibitem{Kresse1996}
G.~Kresse, J.~Furthm{\"{u}}ller, {Efficiency of ab-initio total energy
  calculations for metals and semiconductors using a plane-wave basis set},
  Comput. Mater. Sci. 6~(1) (1996) 15--50.
\newblock \href {https://doi.org/10.1016/0927-0256(96)00008-0}
  {\path{doi:10.1016/0927-0256(96)00008-0}}.

\bibitem{Kresse1999}
G.~Kresse, D.~Joubert, {From ultrasoft pseudopotentials to the projector
  augmented-wave method}, Phys. Rev. B 59~(3) (1999) 1758--1775.
\newblock \href {https://doi.org/10.1103/PhysRevB.59.1758}
  {\path{doi:10.1103/PhysRevB.59.1758}}.

\bibitem{Perdew2008}
J.~P. Perdew, A.~Ruzsinszky, G.~I. Csonka, O.~A. Vydrov, G.~E. Scuseria, L.~A.
  Constantin, X.~Zhou, K.~Burke, {Restoring the Density-Gradient Expansion for
  Exchange in Solids and Surfaces}, Phys. Rev. Lett. 100~(13) (2008) 136406.
\newblock \href {https://doi.org/10.1103/PhysRevLett.100.136406}
  {\path{doi:10.1103/PhysRevLett.100.136406}}.

\bibitem{Grimme2010}
S.~Grimme, J.~Antony, S.~Ehrlich, H.~Krieg, {A consistent and accurate ab
  initio parametrization of density functional dispersion correction (DFT-D)
  for the 94 elements H-Pu}, J. Chem. Phys. 132~(15) (2010) 154104.
\newblock \href {https://doi.org/10.1063/1.3382344}
  {\path{doi:10.1063/1.3382344}}.

\bibitem{Grimme2011}
S.~Grimme, S.~Ehrlich, L.~Goerigk, Effect of the damping function in dispersion
  corrected density functional theory, J. Comput. Chem. 32~(7) (2011)
  1456--1465.
\newblock \href {https://doi.org/10.1002/jcc.21759}
  {\path{doi:10.1002/jcc.21759}}.

\bibitem{Blochl1994}
P.~E. Bl{\"{o}}chl, {Projector augmented-wave method}, Phys. Rev. B 50~(24)
  (1994) 17953.
\newblock \href {https://doi.org/10.1103/PhysRevB.50.17953}
  {\path{doi:10.1103/PhysRevB.50.17953}}.

\bibitem{Anisimov1991}
V.~I. Anisimov, J.~Zaanen, O.~K. Andersen, {Band theory and Mott insulators:
  Hubbard \emph{U} instead of Stoner \emph{I}}, Phys. Rev. B 44~(3) (1991)
  943--954.
\newblock \href {https://doi.org/10.1103/PhysRevB.44.943}
  {\path{doi:10.1103/PhysRevB.44.943}}.

\bibitem{Dudarev1998}
S.~Dudarev, G.~Botton, {Electron-energy-loss spectra and the structural
  stability of nickel oxide: An LSDA+U study}, Phys. Rev. B 57~(3) (1998)
  1505--1509.
\newblock \href {https://doi.org/10.1103/PhysRevB.57.1505}
  {\path{doi:10.1103/PhysRevB.57.1505}}.

\bibitem{Bengone2000}
O.~Bengone, M.~Alouani, P.~Bl\"ochl, J.~Hugel, {Implementation of the projector
  augmented-wave LDA+U method: Application to the electronic structure of NiO},
  Phys. Rev. B 62 (2000) 16392--16401.
\newblock \href {https://doi.org/10.1103/PhysRevB.62.16392}
  {\path{doi:10.1103/PhysRevB.62.16392}}.

\bibitem{Sundman2015}
B.~Sundman, U.~R. Kattner, M.~Palumbo, S.~G. Fries, {OpenCalphad - a free
  thermodynamic software}, Integr. Mater. Manuf. Innov. 4 (2015) 1--15.
\newblock \href {https://doi.org/10.1186/s40192-014-0029-1}
  {\path{doi:10.1186/s40192-014-0029-1}}.

\bibitem{Jain2013}
A.~Jain, S.~P. Ong, G.~Hautier, W.~Chen, W.~D. Richards, S.~Dacek, S.~Cholia,
  D.~Gunter, D.~Skinner, G.~Ceder, K.~A. Persson, {Commentary: The Materials
  Project: A materials genome approach to accelerating materials innovation},
  APL Mater. 1 (2013) 011002.
\newblock \href {https://doi.org/10.1063/1.4812323}
  {\path{doi:10.1063/1.4812323}}.

\bibitem{Joubert2010}
J.~M. Joubert, {A Calphad-type equation of state for hydrogen gas and its
  application to the assessment of Rh-H system}, Int. J. Hydrog. Energy 35~(5)
  (2010) 2104--2111.
\newblock \href {https://doi.org/10.1016/j.ijhydene.2010.01.006}
  {\path{doi:10.1016/j.ijhydene.2010.01.006}}.

\bibitem{Dinsdale1991}
A.~T. Dinsdale, {SGTE data for pure elements}, Calphad 15~(4) (1991) 317--425.
\newblock \href {https://doi.org/10.1016/0364-5916(91)90030-N}
  {\path{doi:10.1016/0364-5916(91)90030-N}}.

\bibitem{NationalInstituteofStandardsandTechnology1998}
M.~Chase, {NIST-JANAF Thermochemical Tables, Monograph 9 (Part I and Part II)},
  4th Edition, Published by the American Chemical Society and American
  Institute of Physics, 1998.
\newblock \href {https://doi.org/10.18434/T42S31} {\path{doi:10.18434/T42S31}}.

\bibitem{Redlich1948}
O.~Redlich, A.~T. Kister, {Algebraic Representation of Thermodynamic Properties
  and the Classification of Solutions}, Ind. Eng. Chem. Res. 40~(2) (1948)
  345--348.
\newblock \href {https://doi.org/10.1021/ie50458a036}
  {\path{doi:10.1021/ie50458a036}}.

\bibitem{Santhy2021}
K.~Santhy, K.~C. {Hari Kumar}, {Thermodynamic modelling of magnetic laves phase
  in Fe–Ti system using first principle method}, Intermetallics 128 (2021)
  106978.
\newblock \href {https://doi.org/10.1016/j.intermet.2020.106978}
  {\path{doi:10.1016/j.intermet.2020.106978}}.

\bibitem{Wenzl1980a}
H.~Wenzl, S.~Pietz, {The Specific Heat Of Iron Titanium Hydride At Room
  Temperature}, Solid State Comm. 33 (1980) 1163--1165.
\newblock \href {https://doi.org/https://doi.org/10.1016/0038-1098(80)90781-4}
  {\path{doi:https://doi.org/10.1016/0038-1098(80)90781-4}}.

\bibitem{Heo2019}
T.~W. Heo, K.~B. Colas, A.~T. Motta, L.~Q. Chen, {A phase-field model for
  hydride formation in polycrystalline metals: Application to $\delta$-hydride
  in zirconium alloys}, Acta Mater. 181 (2019) 262--277.
\newblock \href {https://doi.org/10.1016/j.actamat.2019.09.047}
  {\path{doi:10.1016/j.actamat.2019.09.047}}.

\end{thebibliography}

\end{document}